\documentclass[longauth]{aa}  
\usepackage[utf8]{inputenc}
\usepackage{makecell} 
\usepackage{graphicx}
\usepackage{txfonts}

\usepackage[]{hyperref}
\usepackage{subcaption}         

\usepackage{lscape}

\usepackage{placeins}

\usepackage{multirow}
\usepackage{orcidlink}

\usepackage{multirow}
\newcommand{\espresso}{ESPRESSO}

\newcommand{\sbart}{\texttt{s-BART}}

\newcommand{\epsIndi}{$\epsilon$ Indi}

\newcommand{\metersecond}{${\rm m}\ {\rm s}^{-1}$}

\newcommand{\centimetersecond}{${\rm cm}\ {\rm s}^{-1}$}
\newcommand{\milimitersecond}{${\rm mm}\ {\rm s}^{-1}$}

\usepackage{orcidlink}
\usepackage{float}
\begin{document}

\title{The impact of interpolation in high-resolution spectroscopy}
\subtitle{The overlooked role of interpolation in radial velocity extraction}

\author{A. M. Silva\inst{\ref{inst1}, \ref{inst2}} \orcidlink{0000-0003-4920-738X}
	\and D. Doshi \inst{\ref{inst20}, \ref{inst21}} \orcidlink{0000-0003-3610-3434}
	\and K. Al Moulla\inst{\ref{inst1}}\thanks{SNSF Postdoctoral Fellow}  \orcidlink{0000-0002-3212-5778}
	\and E. A. S. Cristo \inst{\ref{inst1}}                    
	\and É. Artigau     \inst{\ref{inst6}, \ref{inst7}}			\orcidlink{0000-0003-3506-5667}
	\and P. T. P. Viana \inst{\ref{inst1}, \ref{inst2}}   \orcidlink{0000-0003-1572-8531}
	\and N. C. Santos \inst{\ref{inst1}, \ref{inst2}}
	\and J. H. C. Martins \inst{\ref{inst1}}
	\and C. M. J. Marques \inst{\ref{inst1}, \ref{inst2}}  \orcidlink{0000-0002-8784-1448}
	\and S. G. Sousa \inst{\ref{inst1}, \ref{inst2}} \orcidlink{0000-0001-9047-2965}
	\and C. San	Nicolas Martinez \inst{\ref{inst1}, \ref{inst2}} \orcidlink{0009-0003-0571-0554}
	\and T. L. Campante \inst{\ref{inst1}, \ref{inst2}}   \orcidlink{0000-0002-4588-5389}
	\and A. Cabral \inst{\ref{inst3}, \ref{inst15}}  \orcidlink{0000-0002-9433-871X}
	\and S. Cristiani  \inst{\ref{inst10}} \orcidlink{0000-0002-2115-5234}
	}

\institute{
	Instituto de Astrof\'{\i}sica e Ci\^encias do Espa\c{c}o, CAUP, Universidade do Porto, Rua das Estrelas, 4150-762 Porto, Portugal \label{inst1}
	\and Departamento de F\'{\i}sica e Astronomia, Faculdade de Ci\^encias, Universidade do Porto, Rua do Campo Alegre, 4169-007 Porto, Portugal \label{inst2} 
	\and Instituto de Astrof\'{\i}sica e Ci\^encias do Espa\c{c}o, Universidade de Lisboa, Campo Grande, 1749-016 Lisboa, Portugal\label{inst3}
	\and Institut Trottier de recherche sur les exoplanètes, D\'epartement de Physique, Universit\'e de Montr\'eal, Montr\'eal, Qu\'ebec, Canada \label{inst6}
	\and  Observatoire du Mont-M\'egantic, Universit\'e de Montr\'eal, Montr\'eal H3C 3J7, Canada\label{inst7}
	\and INAF-Astronomical Observatory, via Tiepolo 11, 34143 Trieste, Italy \label{inst10}
	\and Departamento de F\'{\i}sica, Faculdade de Ciências, Universidade de Lisboa, Campo Grande, P-1749-016 Lisboa, Portugal \label{inst15}
	\and Department of Physics, McGill University, 3600 Rue University,  Montr\'eal, QC H3A 2T8, Canada \label{inst20}
	\and Trottier Space Institute, McGill University, 3550 Rue University,  Montr\'eal, QC H3A 2A7, Canada\label{inst21}
	}

\date{Received September 30, 20XX}


\abstract
{}
{We explore the impact that spectral interpolation introduces in radial velocity (RV) time-series that are extracted through template-based methods, in particular systematic biases induced in this process.}
{ We generate synthetic datasets with Gaussian profiles to evaluate the flux residuals and line asymmetry that are a result from changing the sampling location of the lines, as induced by barycentric motion and instrumental drifts. We generate synthetic spectra as a sum of Gaussian functions whose parameters were determined through an observed spectrum. The \sbart{} pipeline was applied to such datasets, allowing to evaluate any biases in RV extraction that are introduced by its internal, and implicit, assumptions in line shape. Lastly, we apply the \sbart{} pipeline to \espresso{} observations of four stars, with different observation strategies: two that use high-cadence observations over a single night, and two that have observations spread over multiple nights. When extracting RVs from stellar spectra, we change the interpolation algorithm, used in the process of constructing the stellar template and, afterwards, during RV extraction. They are then compared with RVs extracted whilst using the widely-used cubic-spline interpolation.}
{We find that the synthetic datasets reveal systematic biases with the largest peak-to-peak amplitudes reaching $\sim$ 20 \metersecond{} in low SNR cases, with the amplitude decreasing as the SNR of the spectra increases. In the extreme case of noise-free data, we still recover a systematic bias, albeit at the \milimitersecond{} level, significantly smaller than the RV precision of state-of-the-art instruments. When using real observations, we find that RV time-series that use high-cadence observations with small BERV variation (barycentric motion comparable to the pixel size of the instrument) are impacted by the choice of the interpolation algorithm. This impact is smaller in higher-SNR cases, where the peak-to-peak amplitude reaches $\sim$ 1 \metersecond. In the comparatively lower-SNR case we find  peak-to-peak residuals as large as $\sim$ 25 \metersecond. In cases where the observations are spread over a larger BERV window, we find an upper limit of 20 \centimetersecond{}  of RV scatter for this systematic signal.}
{We conclude that diverse science cases are affected by the results presented in this manuscript, as the interpolation of stellar spectra is present in all of them to place observations in a common wavelength grid. This will impact not only the detection and characterization of exoplanets, but also atmosphere studies, asteroseismic analysis, and even cosmological red-shift determination.}


\maketitle
\nolinenumbers
\section{Introduction}

	The detection and characterization of an Earth-twin stands as a goal of current-day astrophysics, yet to be achieved despite the growth of this exoplanetary field. Such efforts have been the driving factor for the construction of current state-of-the-art spectrographs, e.g. \texttt{EXPRES} \citep{evans_expres_2016}, \espresso{} \citep{pepeESPRESSOVLTOnsky2021}, and \texttt{MAROON-X} \citep{seifahrt_maroon-x_2022}. Nowadays, it is widely accepted that the major roadblock are spurious signals introduced by the host-stars, leading to the development of multiple solar telescopes \citep[e.g., ][]{dumusque_three_2021,lin_observing_2022, farret_jentink_aboras_2022,rubenzahl_staring_2023,santos_poet_2025} with the main goal of understanding stellar activity on the Sun and transposing such knowledge to other systems. Alongside it, we have also seen the development of new mathematical frameworks for the modelling of stellar signals \citep[e.g., ][]{zhaoFIESTAIIDisentangling2022,cameron_separating_2020,aigrain_gaussian_2022,faria_kima_2018}.
	
	The Doppler radial velocity (RV) method is based on the measurement of changes in the motion of a star, along our line of sight. In practice, this is done through the measurement of the displacement of spectral lines in the spectra of the star, through a comparison with their reference positions. Such measurements are often done through the Cross Correlation Function (CCF) method, involving the cross correlation of the stellar spectra with a weighted mask \citep{baranne_elodie_1996,pepe_coralie_2002}. Despite the wide usage of the CCF method, it is not always the optimal approach, especially in the cooler M-dwarf stars. In such cases, data-driven approaches tend to yield better results, either through template matching (TM) algorithms \citep{anglada_escude_HARPS_TERRA_2012,zechmeisterSpectrumRadialVelocity2018,astudillo-defru_search_2015, silvaNovelFrameworkSemiBayesian2022} or line-by-line (LBL) methods \citep{dumusque_measuring_2018,artigau_line-by-line_2022,al_moulla_arve_2025}. These approaches are based on the construction of a stellar template, that will then be compared against the individual observations.

	However, such template-based techniques can still be contaminated by spurious effects during the RV extraction process. \citet{silva_systematic_2025} has recently unveiled a spurious RV signal present in RV time-series, extracted with template-based methods, when all observations are collected within a  short time-frame. In the same vein, \citet{doshi_interpolation_2025} have shown, through simulated spectra, that the interpolation of stellar spectra can also introduce systematic effects in the RV extraction. Such effects are to be expected, as an interpolation of stellar spectra will introduce the implicit assumption of our model being able to correctly represent the shape of a spectral line. If those assumptions do not hold, the shortcoming of our interpolation will play an important role, leading to a possible underestimation of the flux and introducing a systematic bias in the line center, as depicted in Appendix \ref{App:line_center_interpol}. Particularly, in template-based approaches, \citet{doshi_interpolation_2025} has shown that such distortion in the spectral lines are indeed present in the stellar template, biasing the RV extraction.
	
	In this manuscript, we explore the impact of the choice of interpolation algorithms by using both simulated spectra and real observations taken with the \espresso{} spectrograph, both with time-series of short-baseline observations and over a larger time-domain. In Section \ref{Sec:Observations} we describe the dataset that we use, following in Section \ref{Sec:RVmethods} with a description of the RV-extraction tools and the different interpolation algorithms. Section \ref{App:line_shape_interpol} explores the effects of interpolation on the shape of Gaussian profiles, and \ref{Sec:simulations} the RV impact on datasets with purely symmetrical, Gaussian, spectral lines. Then, we follow with an exploration of this impact on real \espresso{} observations on Section \ref{Sec:results}. Lastly, in Section \ref{Sec:conclusions}, we conclude on our analysis and discuss on the impact that such results has on different science cases.
	
\section{Observations} \label{Sec:Observations}
	In order to determine the impact that the interpolation algorithm introduces in stellar spectra, we selected the following targets based on their observation strategies:

	\begin{itemize}
		\item HD40307 - This bright K2 star was observed for 5 consecutive nights in the context of the \espresso{} Guaranteed Time Observations (GTO)\footnote{Observations collected with ESO program 0102.D-0346(A).} with a tentative detection of oscillations \citep{pepeESPRESSOVLTOnsky2021} that were later ruled out by the re-analysis of \citet{campante_expanding_2024}. 
		\item HD209100 (\epsIndi{}) - This K6 star was observed with \espresso{} for 6 consecutive nights on September 2022\footnote{Observations collected with ESO program 109.236P.001} leading to the detection of solar-like oscillations \citep{campante_expanding_2024}. 
		\item Proxima Centauri - This M5 star was also observed in the context of \espresso's GTO program\footnote{GTO observations collected under ESO program IDs: 1102.C-0958(G), 1102.C-0744(I), 1102.C-0744(V), 1102.C-0744(Y), 1102.C-0958(C), 1102.C-0744(M), 106.21M2.001, 1102.C-0744(B), 1104.C-0350(P), 1102.C-0958(N), 108.2254.003, 1102.C-0958(I), 1104.C-0350(K), 1102.C-0744(Z), 1102.C-0744(N), 1104.C-0350(V), 1102.C-0958(A), 1102.C-0744(K), 1102.C-0958(F), 106.21M2.004, 1102.C-0744(A), 106.21M2.006, 1104.C-0350(J), 1104.C-0350(L), 1102.C-0744(T), 1104.C-0350(O), 1102.C-0958(K), 1102.C-0958(H), 1102.C-0744(O), 1102.C-0744(X), 1104.C-0350(M).}, leading to a detection of a candidate sub-earth with an orbital period of 5 days \citep{fariaCandidateShortperiodSubEarth2022}. The star was observed for a period of 3 years, with a total number of 117 observations.
		\item HD39091 - This G0 star was observed in the context of the ESPRESSO GTO\footnote{Observations collected with ESO program IDs  1102.C-0744(A), 1102.C-0744(I), 1102.C-0744(K), 1102.C-0958(G), 1102.C-0958(I), 1102.C-0958(L), 1102.C-0958(E), 1102.C-0958(A), 1102.C-0958(D), 1102.C-0744(B).}, allowing for a precise characterization of the planetary system \citep{damasso_precise_2020}. This system was observed for about half a year, with 204 observations being collected.
	\end{itemize}
	
	This set of stars allows us to cover two different regimes of observations: intensive campaigns for short timespans (HD40307 and \epsIndi) and observing campaigns spread over multiple months (Proxima Centauri and HD39091). From the asteroseismic campaigns we only select for our analysis the first night of each. In Table \ref{App_tab_dset_stats} we present a detailed overview of each dataset.

	\begin{table}[H]
		\caption{Number of points collected for each star, alongside the median cadence, timespan of observations, and the median SNR at $\lambda$=600nm (as measured in ESPRESSO order 120).}              
		\label{App_tab_dset_stats}      
		\centering                                      
		\begin{tabular}{ccccc}\hline \hline
			Star   &  N   & Cadence &  Time span   & $SNR_{600 nm}$  \\ \hline
		  Proxima  & 117  &      9.47 days       & 1098.9 days  &            48.0            \\
		  HD39091  & 204  &     17.35 hours      &  146.7 days  &           230.17           \\ \hline
		  HD40307  & 406  &      78.09 sec       &  8.8 hours   &           78.56            \\
		  \epsIndi{} & 313  &      60.08 sec       &  5.2 hours   &           231.35           \\ \hline
		\end{tabular}
	\end{table}

\section{Methods} \label{Sec:RVmethods}

	\subsection{Radial Velocity extraction}

		In this manuscript we explore the impact of the choice of strategy for spectral interpolation  during RV extraction through a TM algorithm.  However, the underlying effects should also be noticeable in any method that is based on the construction of a data-driven stellar template and subsequent comparison between such a model and individual observations. An example of such is RV extraction through line-by-line approaches, where the individual spectral lines are compared against a template of the line itself and, ultimately, correspond to an outlier-resistant template matching. For our analysis, we use the \textit{Semi-Bayesian Approach for RVs with Template matching} (\sbart{}\footnote{Publicly available at \url{https://github.com/iastro-pt/sBART}}) algorithm \citep{silvaNovelFrameworkSemiBayesian2022}. \sbart{} is a TM algorithm, built around a core assumption: the RV shift that a planetary companion introduces is independent of the wavelength at which it is measured. This is enforced by assuming a common RV shift to describe all differences between the stellar template and the observations with which the template is compared. Its implementation within a semi-Bayesian context also provides a consistent method to characterize the posterior distribution of the RVs. The telluric contamination of the spectra is accounted for through a binary mask, rejecting any lines deeper than 1\%. This is a conservative criterion, that balances the minimization of telluric contamination and preservation of RV content for optical spectrographs. In this manuscript, unless explicitly mentioned differently, the RV extraction was done through this methodology.

		The determination of \sbart{} RVs hinges on the construction of a high-SNR stellar template, that is based on stacking all observations after they are placed in a rest-frame, as defined through a previous estimate of the RV.  For that to be possible, all observations must be defined in the same wavelength grid, which is not necessarily the case due to the way the spectrograph wavelength solution is determined (see Appendix \ref{App:single_mult_differences_linesampling} for a description). As such, all available observations are interpolated to a common wavelength grid that is defined \textit{a priori} and then combined through a weighted mean or equivalent. This wavelength grid can be either taken from one of the observations, fixing the sampling of the template to the one of the data, or can be over-sampled through the interpolation of stellar spectra to a denser wavelength grid. After the stellar template is constructed, the RV displacement between spectra and template is determined through a comparison of their residuals, which must be computed in the wavelength grid of the observation. This leads to a new interpolation of the stellar template so that it can be compared with the selected observation and ultimately leading to the determination of the RV measurement when spectra and template are aligned with one another.

	\subsection{Interpolation of stellar spectra} \label{Sec:interpol_algorithms}
		
		As discussed, when dealing with stellar spectra we often encounter the need to interpolate our observations to a new wavelength grid. In the majority of the cases, this occurs when we want to combine observations to construct a data-driven, high-SNR, model of the stellar spectra\footnote{A stellar template, widely used for RV extraction and in atmospheric characterization studies}. Such interpolations are typically done through a spline-based approach, ensuring that the second derivative of the spectra is defined. However, when introducing an interpolation step to our data analysis recipe, we are implicitly introducing an assumption regarding the shape of the stellar spectra: we assume that both continuum, line wings, and line center can be perfectly modelled by a given parametrization. As we do not have infinite sampling, our spectral lines are discrete realizations of the underlying true stellar spectrum, which will lead to biases in the reconstruction of spectral lines during the interpolation stages (Appendix \ref{App:line_center_interpol} showcases the issue on a Gaussian profile).

		In this manuscript we discuss the impact of the choice of the interpolation algorithm on the construction of an high-SNR stellar template and subsequent application for the extraction of RV time-series in different observation modes. For this purpose, we will use the following algorithms:
		
		\begin{itemize}
			\item Cubic spline: Interpolation of data through a piece wise cubic polynomial, the default interpolation algorithm of \sbart. It will ensure that the derivative is smooth and that the second derivate of the flux is also defined.
			\item Quadratic spline (quad spline): Piece wise interpolation, as implemented by \textit{scipy interp1d} \footnote{\url{https://docs.scipy.org/doc/scipy/reference/generated/scipy.interpolate.interp1d.html}}. Unlike the cubic spline, it does not guarantee a smooth derivative.
			\item Piece wise cubib hermite polynomial (pchip)\footnote{\url{https://docs.scipy.org/doc/scipy/reference/generated/scipy.interpolate.PchipInterpolator.html}}: Uses monotonic cubic splines. The monotonic behavior should be expected between any two pixels in the spectra, seemingly making this suitable for our use case. However, it is important to note that this no longer holds true in the two pixels that surround the center of the spectral lines, with the sole exception of cases where the true center of the line lies in a sampled location.
			\item AKIMA\footnote{\url{https://docs.scipy.org/doc/scipy-1.15.3/reference/generated/scipy.interpolate.Akima1DInterpolator.html}}: Fits piecewise cubic polynomials, leading to a continuously differentiable sub-spline that ensures a smooth interpolated curve.
			\item Floater-Hormann interpolant\footnote{\url{https://docs.scipy.org/doc/scipy/reference/generated/scipy.interpolate.FloaterHormannInterpolator.html}}: Local blend of N polynomials of third degree, working particularly well for equidistant (or near equidistant) nodes. This assumption is held within a given spectral order, allowing us to use this type of interpolation function.
		\end{itemize}

		It is also important to note that \sbart{} ensures that the RV extraction process is only informed by wavelengths that are common to any RV shift of the template, bypassing any extrapolation of stellar spectra. The aforementioned algorithms were added in the \texttt{ASTRA}\footnote{Publicly available at \url{https://github.com/Kamuish/ASTRA}} \citep{silva_astra_2026} package, as implemented within \textit{scipy} \citep{virtanenSciPyFundamentalAlgorithms2020}, so that they can be used by \sbart. 

\section{Impact of interpolation on the shape of spectral lines} \label{App:line_shape_interpol}

	In this Section we evaluate the impact of interpolation of spectral lines through the analysis of Gaussian profiles. First, in Section \ref{App:gaussian_interpol}, we evaluate the impact of interpolation on the flux measurements of the line. Then, in Section \ref{App:line_interpol_flux_residuals} we search for interpolation-induced asymmetries on the shape of the spectral line. 
	
	\subsection{Flux residuals on a gaussian line} \label{App:gaussian_interpol}

		To estimate the impact that the interpolation algorithms introduce in a spectral line, we evaluate its impact on a single gaussian line, whose flux ($\mathcal{F}$) is given by 
		
		\begin{equation} \label{Eq:gauss_line}
			\mathcal{F(\lambda)} = C\cdot(1\ -\ a\ \cdot\ e^{-\frac{(\lambda - \lambda_{c})^{2}}{2\ \cdot\ \sigma^2}}) 
		\end{equation}

		\noindent where \textit{C} represents the continuum level, \textit{a} represents the line depth, $\lambda_c$ the central wavelength of the line, $\sigma$ its standard deviation (i.e., the line width), and $\lambda$ the wavelengths for which the line is defined.

		Using this model, we evaluate the impact of interpolating its fluxes by applying the following recipe:

		\begin{itemize}
			\item We generate synthetic gaussian lines ($Flux_{order}$) using wavelength grids ($\lambda_{order}$) taken from two orders of a \espresso{} file (one in the blue detector, other in the red detector);
			\item We generate an array of possible RV shifts ($\{RV_0, ..., RV_N$\}), spanning between zero and 500 \metersecond{}, a window that covers the mean pixel size in \espresso{};
			\item Each $RV_i$ is used to Doppler-shift $\lambda$ to a new wavelength grid ($\lambda'_{order}$), mimicking the change of sampling that is introduced by the BERV variation over a night (refer to Appendix \ref{App:single_mult_differences_linesampling} for details);
			\item We evaluate the Gaussian model (Eq. \ref{Eq:gauss_line}) on the new grid $\lambda'_{order}$ and interpolate it to the original wavelength grid,  $\lambda_{order}$.
			\item We compute the flux residuals between the original flux  ($Flux_{order}$) and the new one, obtained through the previous step.
		\end{itemize}
		
\begin{figure}[h]
			\centering
			\resizebox{\hsize}{!}{\includegraphics{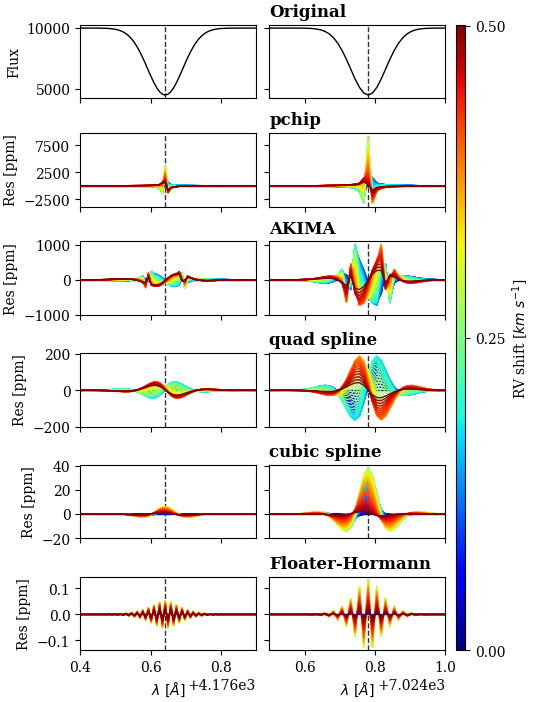}}
			\caption{Flux effect of interpolating a Gaussian line (first row) on the blue (left column) and red (right column) wavelengths of \espresso. The rows after the first one present the residuals between the interpolated gaussian line and the expected one, in parts per million (ppm).}
			\label{Fig:gaussian_impact}
		\end{figure}

		For this preliminary analysis, we fix the depth and width of the Gaussian model to be equal for the two spectral orders. This provides a clear overview of the effect that the wavelength spacing between two pixels has on the residual profile. The central wavelength, $\lambda_c$ is defined to be the mean between two neighboring pixels ($\lambda_c\ =\ (lambda_{i}\ +\ \lambda_{i+1})/2$).	This recipe is applied with all interpolation algorithms that were previously described in Section \ref{Sec:interpol_algorithms}, with the comparison being shown in Figure \ref{Fig:gaussian_impact}. 
		
		From here, we find that the interpolation algorithm that was selected for the interpolation of stellar spectra plays a critical role in the resulting line profile. As one would expect, interpolating data away from the nodes of interpolation -- i.e, between pixels -- leads to higher residuals, as a consequence of the interpolation relying more on our underlying assumptions regarding line shape. In spite of this, we find that for both the cubic and the Floater-Hormann interpolation strategies lead to residuals -- for a single gaussian line -- that are smaller than 100 parts per million. It is also important to note the larger impact on the red detector (right column), introduced by the larger $\Delta\ \lambda$ that exists on this detector and, consequently, a lower sampling of the line.
	
	\subsection{Assymetries introduced on model spectral lines} \label{App:line_interpol_flux_residuals}
		
		As we have seen in Section \ref{App:gaussian_interpol}, the interpolation of a gaussian line translates into strong residuals near the core of the line, with significant differences in the bluer and redder regions of the stellar spectra. As such, it is important to understand how this behavior evolves over the multiple spectral orders and if it introduces any asymmetries in the resulting line profile, than can affect the derived radial velocity. For this effect, we evaluate interpolation-induced asymmetries through the following process:
		
		\begin{enumerate}
			\item We apply \texttt{ARES}\footnote{Publicly available at \url{https://github.com/sousasag/ARES}} \citep{sousa_new_2007, sousa_ares_2015} to one \espresso{} observation of \epsIndi, allowing us to automatically detect all spectral lines and model them through a combination of \textit{N} Gaussian profiles. \texttt{ARES} provides us with the line centers ($\mu_{i}$), depths ($a_{i}$), and widths ($\sigma_{i}$) for all lines that it was capable of detecting.
			\item For each line that was detected, we independently apply the following recipe:
			\begin{itemize}
				\item Construct a wavelength grid, $\lambda$, by selecting a window of 200 points of the \espresso{} wavelength grid, centered around the closest point to the central wavelength of the line. The size of the grid is such that it ensures that the full Gaussian profile fits inside it, without clipping the wings of the line; 
				\item Evaluate a Gaussian function -- with the \texttt{ARES}-fitted parameters -- on $\lambda$, that will serve as the reference profile;
				\item Construct a new wavelength grid, $\lambda'$ by applying a given Doppler-shift to $\lambda$. Then, we re-evaluate the gaussian model in the new grid, without changing the line parameters. This ensures that we do not introduce a change in the central location of the line, only sampling it at a different region.
				\item Interpolate the new Gaussian line from $\lambda'$ to $\lambda$ and compute the flux residuals;
				\item This process is repeated for a vector of Doppler shifts such that we fully sample the mean $\Delta\lambda$, for the selected interval, $\lambda$. This means that the maximum Doppler-shift will vary from the bluer to the redder regions of the spectra;
				\item We sum the residual profiles corresponding to all selected Doppler-shifts and integrate them over the left and right side of the center, providing us with a measure of the asymmetric nature of possible interpolation artifacts.
				\item We also compute the ratio between the line width and the $\Delta\ \lambda$ of the wavelength grid, as well as the difference between the central location of the Gaussian line and the closest point in the wavelength grid.
			\end{itemize}
		\end{enumerate}

\begin{figure}[h]
			\centering
			\resizebox{\hsize}{!}{\includegraphics{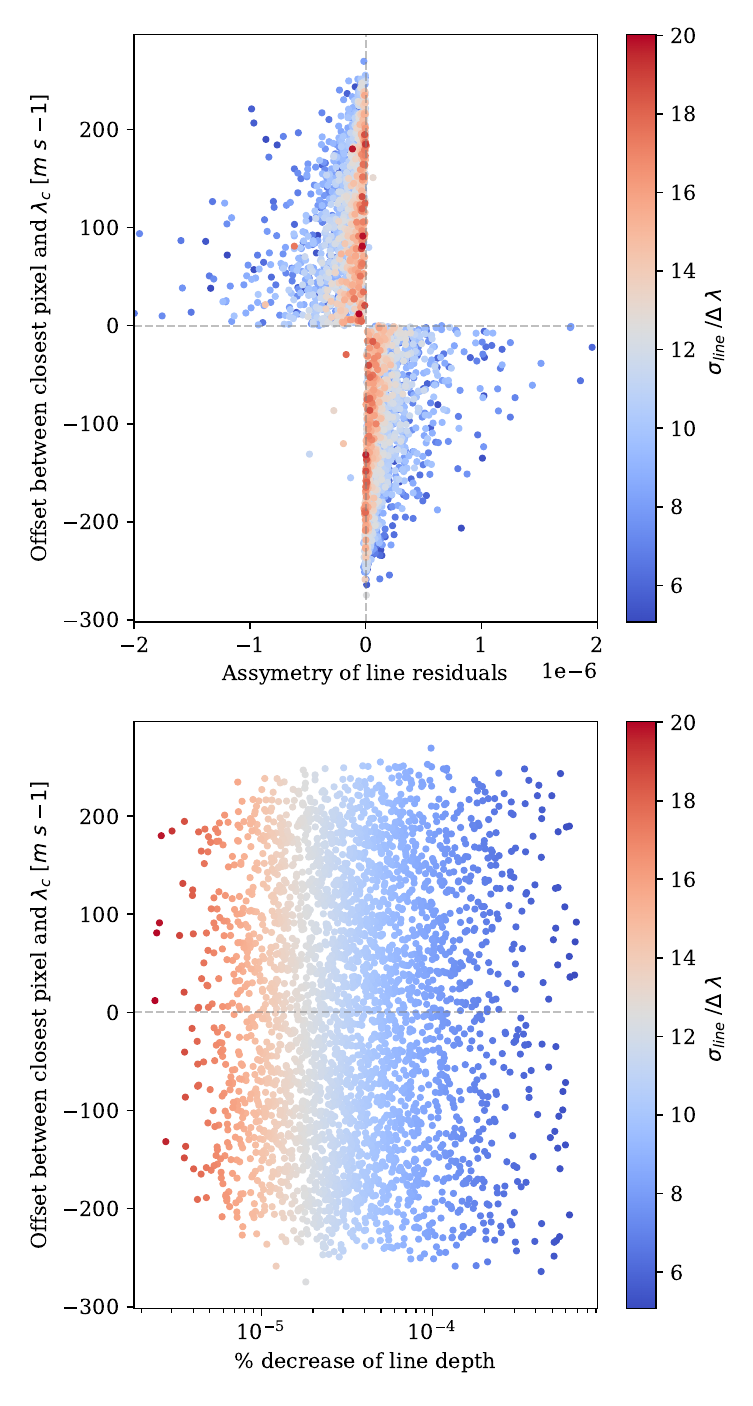}}
			\caption{\textbf{Top:} Asymmetry of the \texttt{ARES}-fitted spectral lines as estimated by integrating the interpolation artifacts when the BERV coverage is enough to fully probe one pixel. \textbf{Bottom:} Median percentage decrease of line depth, computed over the interpolation artifacts when the BERV coverage is enough to fully probe one pixel; The lines, spread over \espresso's wavelength range, are color-coded by the amount of pixels that exist inside its FWHM span.}
			\label{Fig:interpol_gaussian_asymmetry}
		\end{figure}

		In Figure \ref{Fig:interpol_gaussian_asymmetry} we show the result of this analysis, when using a cubic-spline interpolation. We see that when the closest-to-center grid point lies on the left side of the true center (negative values in the vertical axis) we have larger assymetries on the left side of the line (positive values in the horizontal axis). This behavior is mirrored when the closest point is on the right side of the true center.

		If we now focus on the very well sampled lines in our dataset ($FWHM/\Delta\ \lambda\ >$ 14, in dark-blue colors) we unsurprisingly find that the location of the true center of the line, respective to the grid locations, does not introduce a strong asymmetry on the final residual profile. If we focus on the other, less sampled lines, we find that that is no longer the case. As we decrease the number of grid points within a line (transition from green to red colors) we find larger asymmetries. From here, we find that a high spectral sampling is key to decrease interpolation induced asymmetries on the shape of the lines. Focusing on the variation of line depth, in the bottom panel, we also find that the line depth is more affected by interpolation in the less sampled lines. From here, we find that traditional activity indicators, e.g. line emission, can be impacted by such effects. However, it is not clear how such asymmetries will translate into RV biases, as there is no guarantee of where the true center of the Gaussian line lies, relative to the wavelength sampling of our spectrograph.

\section{Impact on simulated observations} \label{Sec:simulations}
	
	In Sect. \ref{App:line_shape_interpol} we found that the interpolation algorithm will introduce an asymmetry in the line profiles, which can then be translated into a RV bias. As such, in this Section we probe the sensitivity of the RV extraction to such effects, under the assumption of perfectly symmetrical line profiles, that are generated through a Gaussian function (Eq. \ref{Eq:gauss_line}). In order to do so, we test two scenarios: one where we have a single spectral line per order, and other where we use all \texttt{ARES}-detected lines. In these simulations we will include a wavelength solution that accounts for the effects of the BERV on the line sampling (refer to Appendix \ref{App:single_mult_differences_linesampling}) without introducing any RV shift on the data. Throughout this Section we fix our interpolation scheme to the widely used cubic-spline interpolation.

	It is however important to note that the symmetric nature of the spectral lines that we consider does not reflect the true nature of stellar spectra, where the lines are often blended and have a more complex shape that evolves over wavelength and time. The specification of a per-line asymmetry parameter would introduce a new degree of freedom in the simulations, as we would need to arbitrarily define it. Furthermore, the asymmetry of the lines is not constant over time, as a consequence of stellar activity. As such, we opted for a simpler model of the spectra, allowing to isolate the effects of interpolation on the RV extraction, without the need to introduce additional assumptions.

	\subsection{RV signal on a single gaussian line} \label{Sec:RV_single_gauss}
		The first step in our analysis was to evaluate this effect on a single Gaussian line, without injecting any noise, i.e., when using a perfect Gaussian line. We start our analysis by constructing a stellar spectrum that only contains a single Gaussian line per spectral order, through the following process:

		\begin{enumerate}
			\item We start by selecting one \espresso{} observation of a given star, from which we will select its wavelength grid to be our reference, $\lambda_{ref}$ that will be used to generate the other synthetic observations. 
			\item Each synthetic observation will be defined at a wavelength $\lambda$ that is obtained by Doppler-shifting $\lambda_{ref}$ by a given RV ($\lambda\ =\ \lambda_{ref}\ \cdot\ (1\ +\ RV/c)$, where $c$ is the speed of light.).
			\item The fluxes of each spectral order will contain a single, Gaussian, spectral line, as  given by Eq. \ref{Eq:gauss_line}.		
		\end{enumerate}

		Using this recipe, we construct 100 observations, corresponding to a vector of RV shifts in the interval $\{0, 500\}$ \metersecond, with a step of 10 \metersecond. This interval allows us to fully probe the mean \espresso{} pixel size in RV (500 \metersecond). The observations are then processed through the usual workflow for RV extraction, with the construction of a stellar template and subsequent alignment between it and the individual observations. It is important to stress that through this procedure we create a synthetic dataset with no intrinsic RV shift -- as the parameter of the Gaussian lines are kept constant -- with only a variation of the sampling location of the lines induced by apparent BERV shifts.  

		In this case, the RV extraction is accomplished through a classical template matching algorithm, allowing us to compute individual RVs per spectral order, which are then combined through a weighted mean to yield a final RV for the observation. This algorithm is also implemented within the \sbart{} pipeline, allowing us to evaluate the systematic signal at different wavelengths. 
		
		\begin{figure}[h]
			\centering
			\resizebox{\hsize}{!}{\includegraphics{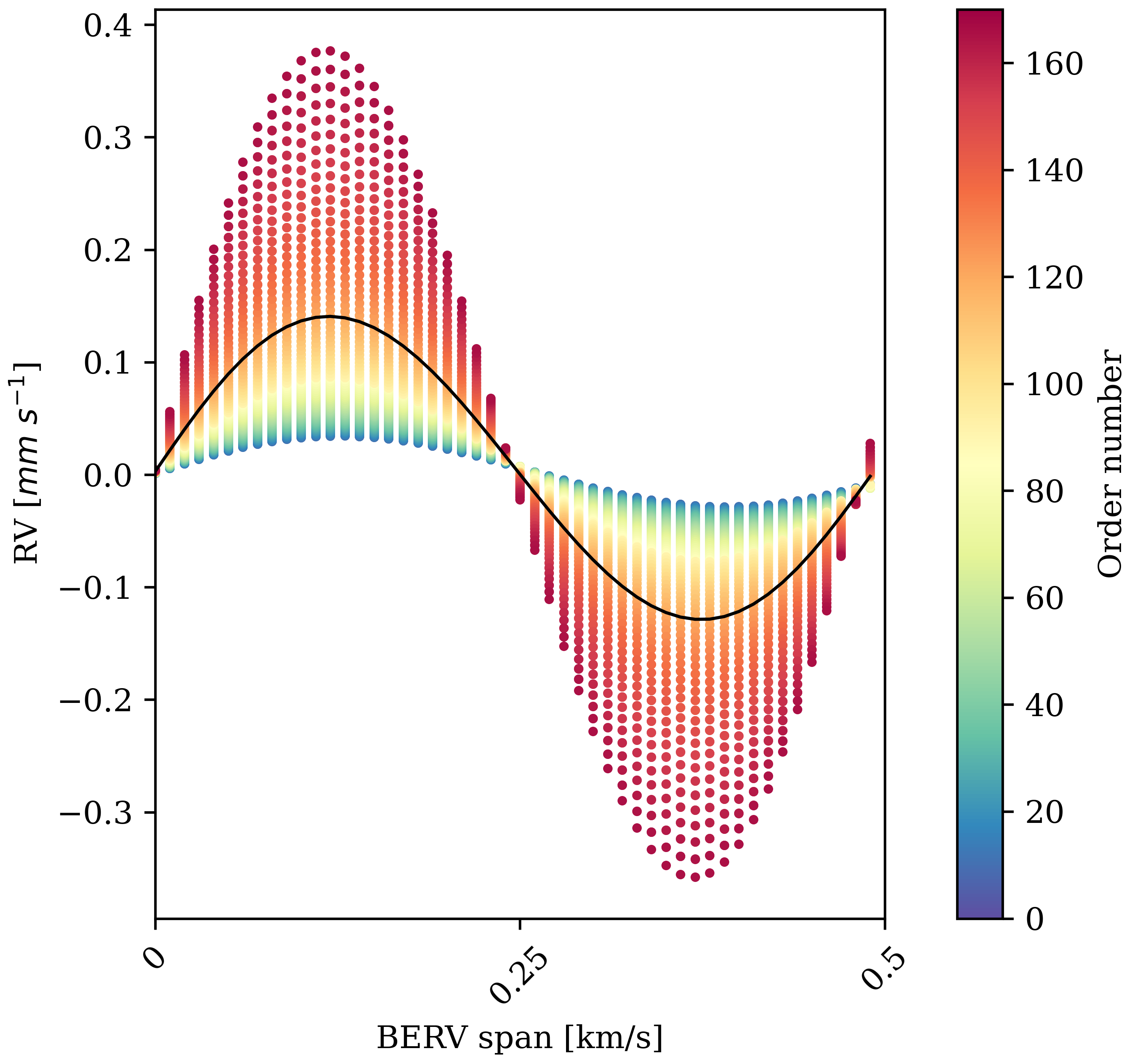}}
			\caption{RV curve of the same gaussian profile placed on different \espresso{} orders, being impacted by a sampling variation from RVs in the interval [0, 500] \metersecond. The spectral orders are color-coded, as per the colorbar, whilst the combination of the order-wise measurements are shown in black.}
			\label{Fig:single_gaussian_rv_signal}
		\end{figure}
		
		The results from this analysis are presented in Fig. \ref{Fig:single_gaussian_rv_signal}, showing that even in a perfect, noise-free, scenario we recover a systematic signal that is introduced by the interpolation of the spectra during the RV extraction. As expected, we find a larger impact in the redder spectral orders, as a consequence of the larger wavelength coverage of each pixel, enhancing the RV impact. Despite this coherent systematic signal, as a function of the BERV excursion, the peak-to-peak amplitude of this signal is at the level of the \milimitersecond{}, significantly smaller than the RV precision of state-of-the-art spectrographs (e.g., \espresso{} has an expected RV precision of 10 \centimetersecond{} \citep{pepeESPRESSOVLTOnsky2021}). Furthermore, the clear chromaticity of this effect will also lead to a smaller systematic signal when the order-wise RVs are combined into a common RV for the full observation, as seen through the black curve. In Appendix \ref{Sect:spectral_resolution} we present the impact that the instrumental resolution introduces in the RV extraction from a noise-free dataset. It is also important to note that the retrieved RV curve crosses a RV of zero both near the edges of the pixel, where we are interpolating closer to the nodes, and in its middle where the interpolation is equally afar from both nodes.
		
\begin{figure}[h!]
			\centering
			\resizebox{\hsize}{!}{\includegraphics{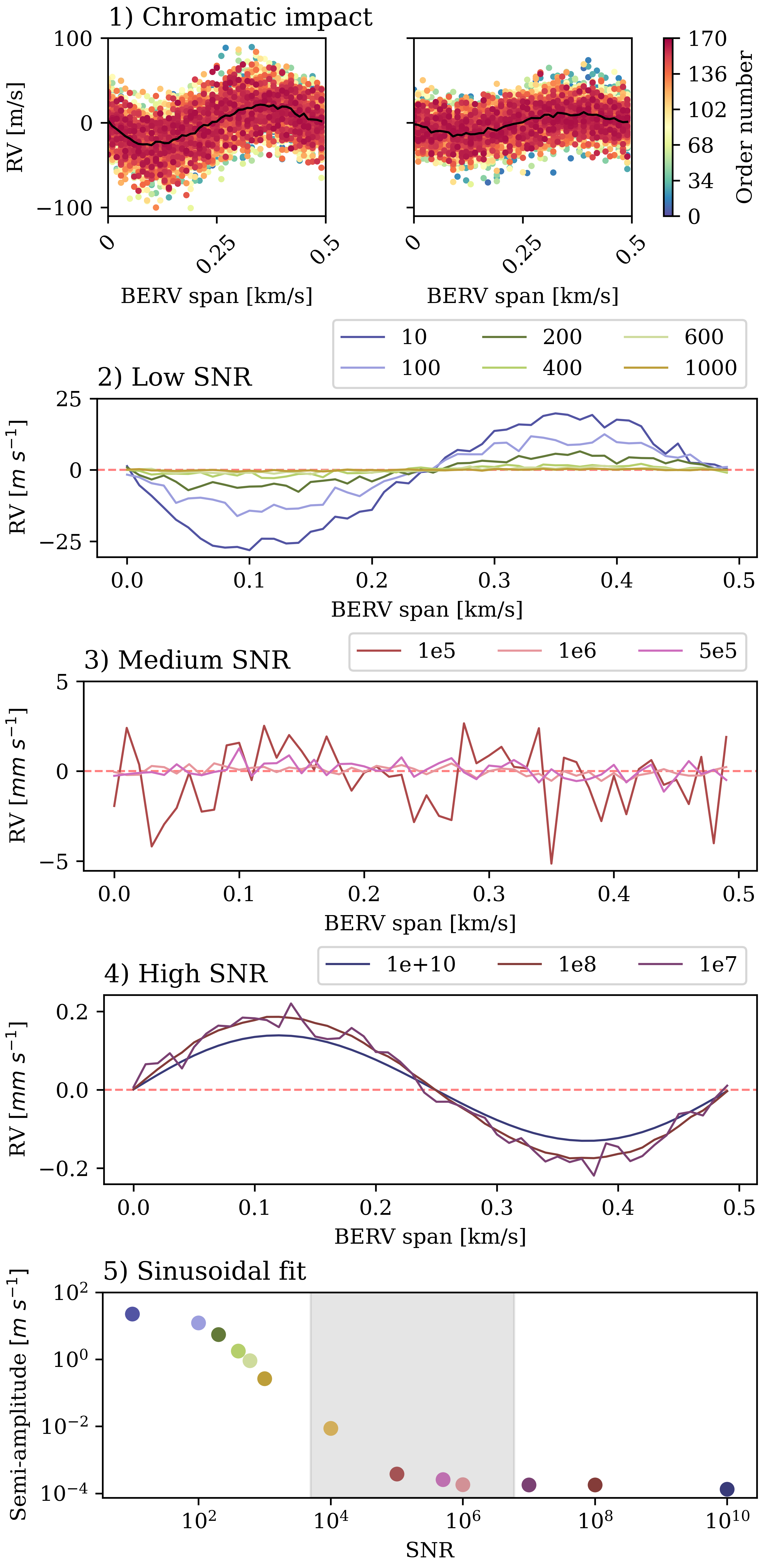}}
			\caption{Impact of adding Poisson noise to the RV curve of the same Gaussian profile placed on different \espresso{} orders, being impacted by a BERV variation in the interval [0, 500] \metersecond. \textbf{Top row:} Chromatic impact for an SNR of 10 (left) and 100 (right), with the combined RV measurement over imposed as the black curve. The spectral orders are color-coded, as per the colorbar. \textbf{2nd to 4th row:} Combined RV measurement for the different SNR levels, as a function of the BERV variation, divided into three regimes of low, medium, and high-SNR. \textbf{5th panel:} Amplitude of a sinusoidal fit to the RVs curves.  The shadowed region represents the SNR regime in which the sinusoidal shape is not present.}
			\label{Fig:single_noisy_gaussian_rv_signal}
		\end{figure}

		Following the results from the noise-free Gaussian model, we repeated the same analysis after injecting Poisson noise in the spectra, so that we can emulate different Signal to Noise Ratios (SNR). The results of this new comparison are shown in Fig. \ref{Fig:single_noisy_gaussian_rv_signal}, where we see a large dependency on the SNR of the retrieved sinusoidal profile, reaching a peak-to-peak amplitude of up to 40 \metersecond{} on a dataset with an SNR of 10. In order to analyze the dependency of the amplitude on the SNR of the data, we fit a sinusoidal model to each RV profile, with this metric shown in the bottom plot of this Figure.
		
		If we start by focusing our attention on the upper panel, we find once again a large  spread of the order-wise radial velocities, especially in the lower SNR case. Similarly to before, the combination of those independent measurements into a common value lead to a peak-to-peak amplitude that is smaller than the one found at the level of the individual spectral orders. In the low SNR cases, we find an impact at the level of multiple \metersecond{}, with its amplitude being inversely correlated with the SNR of the observations. Through the amplitude of the sinusoidal fit to the final RV measurements we find a convergence towards the \milimitersecond{} regime that was previously found in the noise-free scenario (Figure \ref{Fig:single_gaussian_rv_signal}).

		Interestingly, when moving from low-SNR case towards the infinite-SNR case we find three different regimes in the RV: we find that in low- and high-SNR regimes we retrieve a sinusoidal profile in the RV curves, albeit with different phases. In an intermediate SNR-regime (as shadowed in the Figure) we find an almost flat profile, albeit with still decreasing amplitude. The behaviour that is retrieved in the low- and high-SNR regimes matches our expectations: close to the edges of the pixel, the residuals are small due to interpolation closer to the nodes. In the middle of the pixel, one could also expect to find a small impact, as the interpolation occurs equidistantly to both nodes. The phase inversion, retrieved when moving between the two regimes, is not expected. We postulate that it could be introduced by two different systematic effects: i) One that is deterministic in nature, seen in the extremely high-SNR cases, as well as in the noise-free line; ii) A stochastic component that appears due to injected noise, that is present in the lower SNR regimes. This can then explain the difference behavior, as in some cases we have one of the effects dominating (low- and high-SNR) and, at some given SNR level, with a smaller amount of Poisson noise, we have a destructive interference and recover a flat profile. It is, however, important to note that for the extremely high-SNR cases, we are recovering a \milimitersecond{} signal, well below the noise-floor of current spectrographs.

	\subsection{A forest of Gaussian lines} \label{Sec:gauss_forest}
		Following the results of our simulation with a single Gaussian line, we move to a simulation that is closer to a real stellar spectrum, using a realistic number of lines and line parameters. To do so, we now construct  synthetic spectra through a sum of analytical functions that will, once again, be evaluated at different wavelength solutions without changing the central location of the lines, as per the recipe previously outlined in Section \ref{Sec:RV_single_gauss}. In this case, however, we update our model to a sum of Gaussians, whose parameters were previously determined in Section \ref{App:line_interpol_flux_residuals}. Then, we apply the following steps:

		\begin{enumerate}
			\item For each spectral order in our wavelength basis, we search for all Ares-detected lines that fall within it. Within one synthetic spectral order, the fluxes, F, associated with the wavelength solution, $\lambda$ are now given by: 
			\begin{equation}
				F(\lambda) = 1 - \sum_{i\ =\ 0}^{N\ lines} a_i\ \cdot\ e^{- \frac{(\lambda\ -\ \mu_i)^{2}}{2\sigma_i^2}}
			\end{equation}

			\noindent where \textit{i} represents the \textit{i}-th Gaussian with a center in the window $\lambda$. Blended lines were rejected in the previous step to avoid spectral regions with a $F(\lambda)$ smaller than zero for deep blended spectral lines.
			\item In the lower-SNR, and noisier, regions of the spectra we find that ARES can fail to properly detect the stellar continuum in a uniform fashion across neighboring lines, possibly leading to non-physical, negative, flux values. As such, we reject any detected line whose flux would introduce such non-physical values in the spectra.
			\item Then, we assume a given SNR of our data, adding Poisson noise to the synthetic spectra and computing the associated flux uncertainties.
		\end{enumerate}

\begin{figure}[ht]
			\centering
			\resizebox{\hsize}{!}{\includegraphics{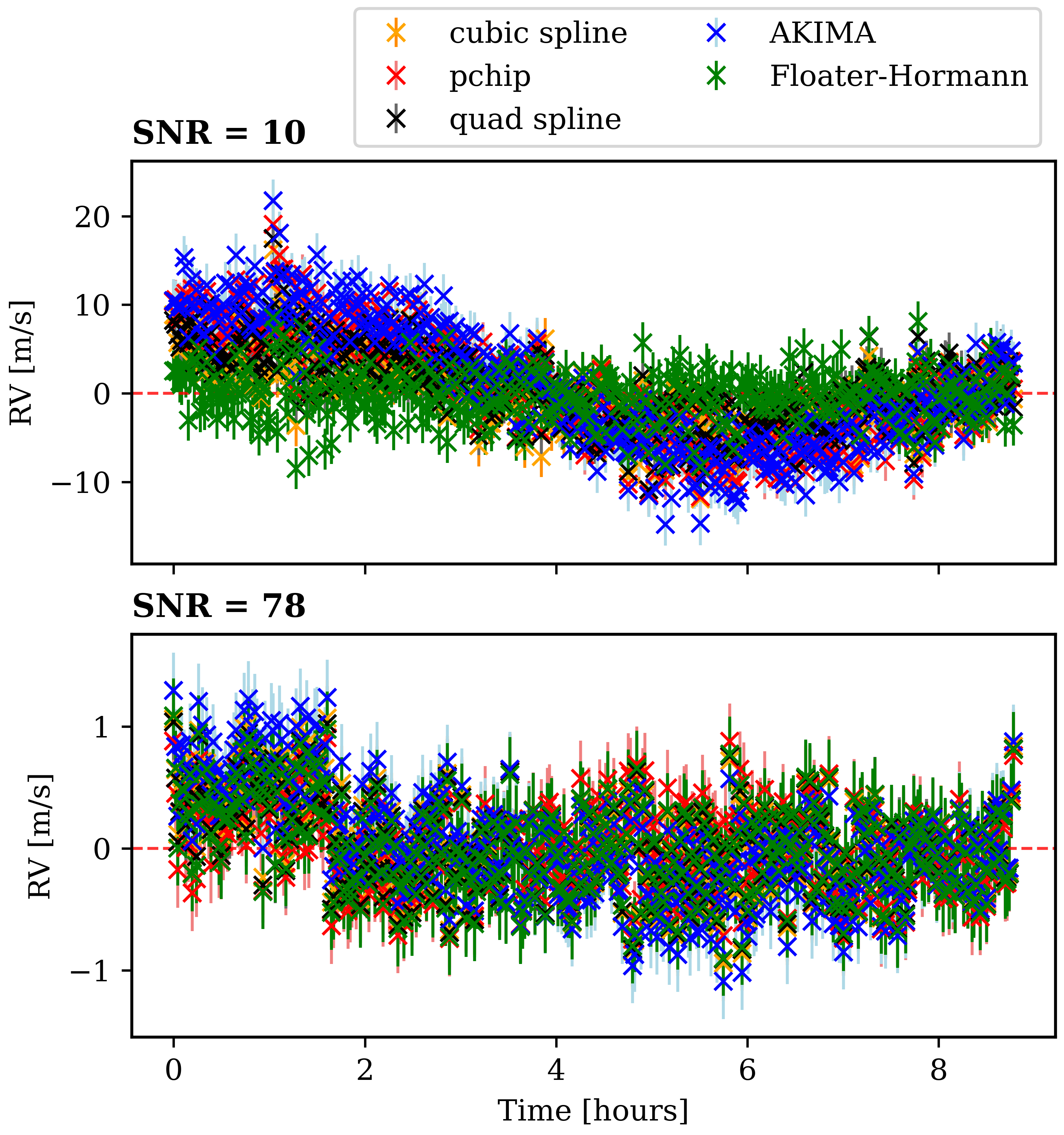}}
			\caption{RV signal from two simulated datasets consisting of a sum of Gaussian profiles, using the BERV profile of one night of HD40307. The different colors represent different interpolation strategies. In the top panel, a flux SNR of 10, and in the bottom one an SNR of 78.}
			\label{Fig:sum_of_gauss_rv}
		\end{figure}

		For this analysis we selected one \espresso{} observation of HD40307 for our wavelength basis, and used the BERV variations of its first night to induce the change of sampling of the analytical profile, leading to a total BERV excursion of 406 \metersecond. The observations are then fed into the \sbart{} pipeline for the extraction of the RVs associated with each individual spectrum. The results of this simulation are shown in Fig. \ref{Fig:sum_of_gauss_rv}, revealing a behavior similar to the one found on a single Gaussian line across the spectral orders. Once again, we retrieve the sinusoidal shape on the RVs, revealing the systematic signal that is induced by artifacts of the underlying interpolation algorithm. An increase of SNR also yields a sharp decrease of its amplitude, with the dataset with an SNR of 78 only presenting some traces of its presence. This value was selected as it is the median SNR level of the \espresso{} observations of HD40307 (refer to Table \ref{App_tab_dset_stats}). Similarly to our previous results, we find a clear impact from the selected interpolation algorithm, leading to larger peak-to-peak RV variations.

		The results from this Section show a clear systematic bias that is introduced at the RV extractions stage, as a consequence of spectral interpolation. Even though we only explore template matching in this manuscript, it is likely that all applications that rely on interpolation of stellar spectra, with small BERV coverage, will be contaminated by such effect. It is also important to note that this sinusoidal signal that we find on the simulated data is not found on real observations. Instead, \citet{silva_systematic_2025} shows the presence of a multi-\metersecond{}, quasi-linear, trend when the BERV coverage is similar to the one in our simulations. We attribute the differences between these two signals to the simplistic nature of our simulations in this analysis: i) We are considering perfectly symmetrical, gaussian, spectral lines, which will not be the case in real observations; ii) Our simulations assume a fixed SNR both intra- and inter-orders, which does not comply with both the instrumental profiles nor the SED of the chosen target star; iii) Our simulations assume a flat continuum, with no variations over time; iv) We do not consider any detector-frame contamination, such as leftover tellurics. Despite the shortcomings of our simulations, we note that the peak-to-peak amplitude lies close to the amplitudes reported in \citet{silva_systematic_2025}.
	
\section{Impact of the interpolation algorithm on ESPRESSO observations} \label{Sec:results}
	
	Our results have shown that the interpolation of stellar spectra introduces artifacts in the resulting profile, introducing \metersecond{} signals on simplistic gaussian datasets. It is likely that such biases are also present in our real observations, affecting the extracted RV time-series. In real observations however, we do not know the true underlying spectra nor RV and, consequently, we must instead evaluate the impact that a change of interpolation algorithm has on the RV curves. In this Section we explore the impact that the interpolation algorithm introduces in the RV extraction with two different observing strategies: i) Collection of observations occurs within a single night of observations; ii) Observations collected over multiple months and years.

	\subsection{Impact on single-night observations} \label{Sec:results_intra}
		
		We start our analysis with the single-night asteroseismology targets (HD40307 and \epsIndi). It is also important to highlight that this type of observation strategy is the one in which a systematic quasi-linear trend appears in RV time-series extracted from template-based approaches \citep{silva_systematic_2025}.
				
		For this comparison we compute the difference between the RVs extracted when using different algorithms to interpolate the stellar spectra (refer to Section \ref{Sec:interpol_algorithms}) and those  when using the default \sbart{} interpolation strategy (cubic spline). This analysis is done when extracting the RVs from the full stellar spectra and from a chromatic sub-division -- we compute independent RV time-series for the blue and red detectors of \espresso{}\footnote{The blue detector spans the first 90 slices of the spectra (380-525nm), whilst the red one spans slice 90 to 170 (525-788nm).}.		
		
\begin{figure*}[h!]
			\centering
			\includegraphics[width=17cm]{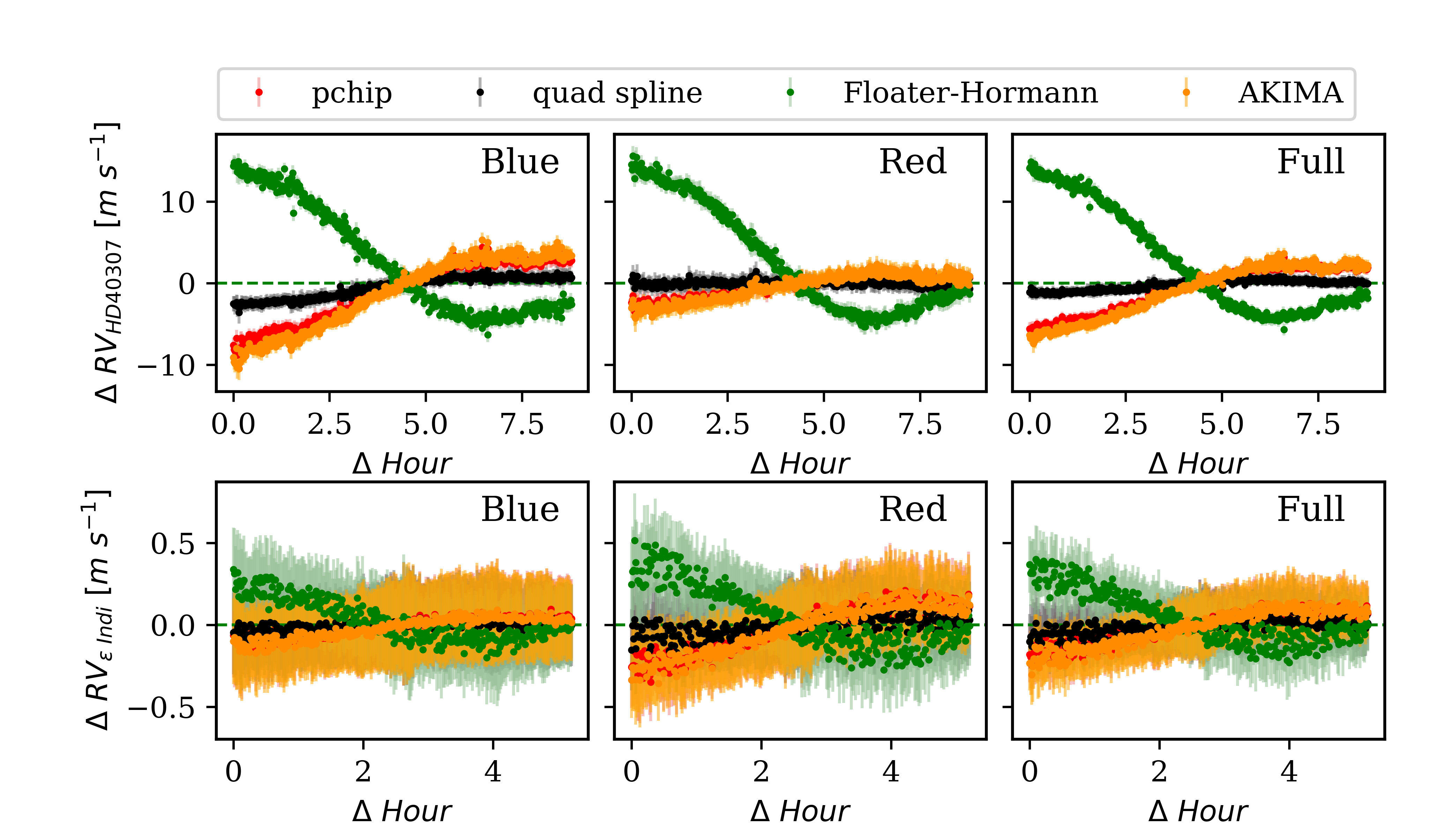}
			\caption{Residuals between \sbart{} RVs with a cubic spline interpolation (the default one) and \sbart{} RVs when using different interpolations. The top row presents the comparison for the low-SNR HD40307 start, whilst the bottom one presents the data of \epsIndi. This analysis was carried out using only the blue (left column) and red (middle column) detectors, as well as using the full spectra (right column).}
			\label{Fig:single_night_interpol_impact}
		\end{figure*}

		The results of such comparison are presented in Figure \ref{Fig:single_night_interpol_impact}, where we find different behaviors for the two targets:
		
		\begin{itemize}
			\item In the lower SNR star (HD40307) we find that the selection of the interpolation algorithm introduces a significant difference in the RV signal that we retrieve, leading to peak-to-peak residuals of the order of 15 \metersecond{}. Interestingly, we find different behaviors depending on the algorithm that we use, with the Floater-Hormann interpolation leading to a larger RV trend than the current cubic-spline interpolation. The other interpolations lead to a decrease of its magnitude, with larger effects on the blue detector than on the red detector.
			\item In the higher-SNR case (\epsIndi) this is no longer the case. Despite finding differences to the cubic-spline RVs, all approaches lead to RVs that agree at the 1-$\sigma$ level, with peak-to-peak differences of the order of the meter per second. 
		\end{itemize}
		
		Previously, we found in our simulations that the \textit{Floater-Hormann} algorithm lead to the smaller residuals, even in the cases where the SNR of the observations is small. However, in real observations we find that this algorithm leads to an increased systematic signal, as seen in Figure \ref{Fig:single_night_interpol_impact}. It is likely that this blend of N polynomial yields the optimal performance under the symmetrical nature of the simulated spectra, which does not match real observations. This analysis makes it clear that the choice of the interpolation algorithm plays a significant role in the RV signals that we are retrieving, with the magnitude of the impact being clearly correlated with the SNR of the target. It is however important to stress that this analysis  does not allow concluding on which algorithm yields the best results, but it does allow for an evaluation of the impact of the assumptions regarding the shape of the stellar spectra imposed by the interpolation algorithm.

	\subsection{Impact on multi-night observations}  \label{Sec:results_multi}
		
\begin{figure}[H]
			\centering
			\resizebox{\hsize}{!}{\includegraphics{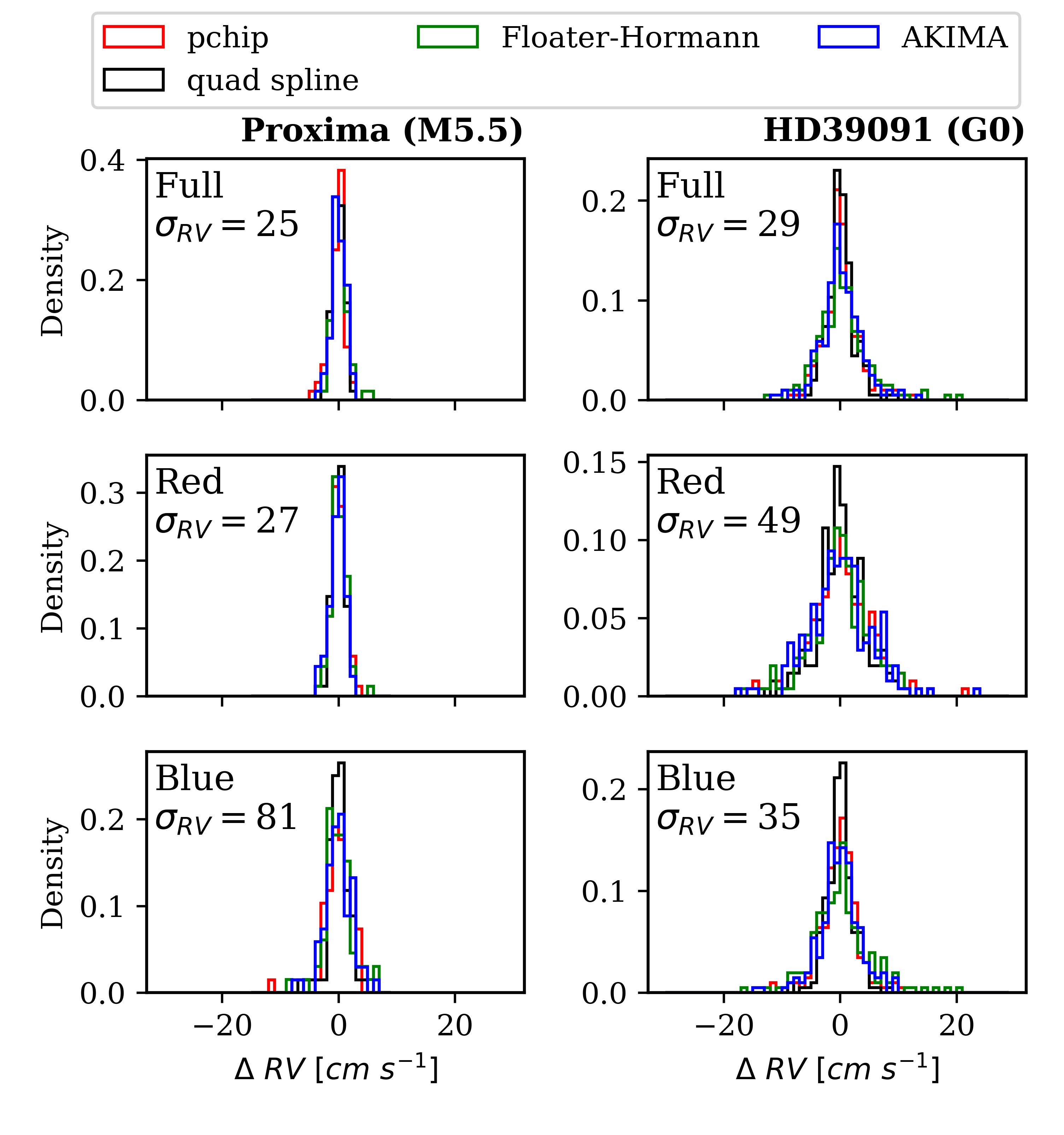}}
			\caption{RV residuals between the \sbart{} RVs obtained with a cubic-spline interpolation (default) and other interpolation algorithms. The analysis is done for two stars, Proxima Centauri (left column) and HD39091 (right column). We present RV time-series when using the full spectrum (top row), and only the red and blue detectors (second and third row, respectively). In each panel we present the median RV uncertainty, in \centimetersecond, in the time-series extracted with a cubic spline interpolation.}
			\label{Fig:multi_night_interpol_impact}
		\end{figure}

		Following the results of Section \ref{Sec:results_intra}, it is important to understand if the interpolation of stellar spectra is impacting the radial velocity time-series that are typically used for the detection and characterization of exoplanets, i.e., those that are collected over multiple nights spread over a period of months or years. We repeat the same exercise as before, now on \espresso{} observations of two different stars, presenting the results in Figure \ref{Fig:multi_night_interpol_impact}.

		We find in both cases that the algorithm selected for the interpolation of stellar spectra imprints variability in the measured RVs, with peak-to-peak amplitudes smaller than 20 \centimetersecond{}. A comparison with the median RV uncertainty, as presented in each panel, shows that the differences are well below the 1-$\sigma$ level. Furthermore, we do not find any systematic differences between the two detectors. The fact that we find a clear impact of the interpolation strategy on single-night observations could be explained through the physical effect that the BERV variation introduces in the in-detector location of spectral lines (refer to Appendix \ref{App:single_mult_differences_linesampling} for a complete description). In multi-night observations, we no longer find a smooth variation of the BERV over the night, thus breaking a possible systematic flux contamination that is then translated to RV variability. It is however important to note that this analysis does not allow us to compute the true impact at the level of the RVs, as those are unknown to us. Instead, this value should be taken as an upper limit of the RV contamination introduced by a change in the interpolation scheme.

\section{Conclusions and discussion} \label{Sec:conclusions}

	In this manuscript we explored how spectral interpolation schemes used during RV extraction modify the shape of spectral lines and introduce RV biases. For this purpose, we applied the \sbart{} RV extraction pipeline to both simulated and real \espresso{} observations, evaluating the impact of different interpolation algorithms on the resulting line profiles and RV time-series. In the context of RV extraction through template-based methods, flux interpolation occurs both at the level of the construction of the stellar template and during its alignment to the individual observations. 
	
	We first looked at line-shape asymetries through simulated datasets consisting in Gaussian profiles, discretized on the wavelength grid of \espresso, whose parameters where retrieved through the \texttt{ARES} pipeline. Our tests with purely Gaussian profiles reveal a clear residual pattern when the interpolation algorithm fails to model the intrinsic stellar spectra within a single pixel. We find that for the widely used cubic spline interpolation, the interpolation-driven residuals can reach a peak-to-peak amplitude of 60 parts-per-million near the core of in a single line. Such residuals were present for the different interpolation methods that we tested, with the majority of them introducing a larger impact close to the line core. Not only do we find a flux residual, but also a clear asymmetry in line shape that is highly dependent of the sampling location of the line and how many pixels it covers. We find a clear asymmetrical pattern on both wings of the spectral line, correlated with the relative position of the true central location of the line and the closest pixel in the wavelength grid. Lastly, we have also shown that an increased wavelength sampling, relative to width of the line, leads to a smaller asymmetry between the left and right side of the wing.
	
	This systematic bias of the stellar flux due to interpolation also imprints a pattern into RV time-series extracted in simulated datasets with no injected Doppler shifts. We find that in low SNR regimes (SNR = 10) we retrieve a bias with a maximum peak-to-peak amplitude of $\sim$ 20 \metersecond{}, consequence of only the change of the wavelength sampling locations of the spectra due to the barycentric motion. This signal is anti-correlated with the SNR of our observations, but it is also retrieved in a noise-free regime, i.e. purely Gaussian profiles with no injected Poissonian noise. However, the retrieved signal is at the level of the \milimitersecond{}, well below the RV precision of the current generation of state-of-the-art spectrographs. If we compare these results with the ones from \citet{silva_systematic_2025} we find similarities: the reported RV systematic effect was inversely correlated with the SNR, i.e., larger signals for smaller SNRs. The main difference comes in the shape of the signal that we retrieve, as the real observations of \citet{silva_systematic_2025} revealed a quasi-linear trend, whilst our simulations yield a sinusoidal signal. We postulate that this difference is introduced by the simplistic nature of our simulations, as they are not comparable to real observations.
	
	Our evaluation of different interpolation algorithms on real \espresso{} observations, spanning a single night, show a striking amplitude of residuals, with the same dataset presenting up to 10 \metersecond{} difference when altering the interpolation algorithm. In the results of this manuscript, we find that the SNR plays a significant role in the RV residuals that we retrieve, with the choice of interpolation algorithm playing a bigger factor in the noisier dataset. This can be interpreted as the noisier spectra (lower SNR) having more inflection points due to noise, consequently increasing the number of locations in which the assumptions of the interpolation plays a more significant role.	Once again, this shows that our current methods for the interpolation of stellar spectra are not capable of properly modelling the spectral lines when we have a small BERV excursion, e.g. BERV variations comparable to the pixel size. It is thus very likely that interpolation induced artifacts can explain a part of the quasi-linear trend reported by \citet{silva_systematic_2025}, despite the difference in the overall systematic profile. As our simulations do not reveal the same amplitude of signals at comparable SNR values (e.g., at a SNR of 78 we no longer find a \metersecond{} contamination in the simulations) it is difficult to support that the quasi-linear trend is purely introduced by interpolation artifacts. 
	
	Our analysis of observations spread over a larger BERV window also reveal some differences in the recovered RVs, albeit with a contamination upper limit of 20 \centimetersecond{} on \espresso, which is at the level of its expected RV precision. It is likely that the interpolation algorithm in this case is similarly introducing an RV bias, albeit just under the current noise floors of our instruments. It is also important to note that there results are well in agreement with the analysis of \citet{doshi_interpolation_2025}, where the RV scatter decreased with the increase of the wavelength sampling, alongside the placement of a noise floor at 0.5 \metersecond{}, for M dwarfs observed with SPIRou. For an increased line sampling, similar to the one of \espresso, such a noise floor falls to the level of $\sim$ 3 \centimetersecond{} in the noise-free simulations. As we have seen throughout this manuscript, the addition of noise to the simulations leads to an increase RV effect, explaining the differences between our observations and the simulations of  \citet{doshi_interpolation_2025}. Lastly, it is also important to note that the underlying methodology of template construction will also play a role. Depending on the selection of the wavelength grid, one can preserve the wavelength spacing of the observations or super-sample them. 
	
	Despite this relatively small contamination ceiling, it can still influence diverse science cases, that are heavily dependent on ultra-precise and stable RV time-series over timespans of multiple years, even when its presence is not straightforward. One example of such, would be the construction of the typical one dimensional spectra (S1D) from the stitching of multiple spectral orders (S2D). The resampling of the two-dimensional data onto a common wavelength grid can be accomplished through an interpolation stage, thus contaminating such data products. Similarly, the determination of the instrumental drift, through a comparison of Fabry-Pérot (FP) calibration exposures, simultaneous to science-observations, to a reference frame can also be affected \citep[e.g. if employed through TM-based approaches][]{bauer_carmenes_2020}
	
	Summarizing our results, it follows that other spectra-based science cases will also be impacted by the same systematic effects on the flux, as long as they use high-cadence observations taken during a short period of time:
	
	\begin{itemize}
		\item The characterization of exoplanetary atmospheres via transmission spectroscopy critically depends on the construction of high-SNR stellar templates. These templates serve as reference spectra for comparing in-transit observations against an estimate of the stellar spectrum behind the planet. In practice, they are typically built from out-of-transit observations, which are shifted into the stellar rest frame and co-added to maximize signal-to-noise, thereby enhancing the detectability of atmospheric absorption features in the planet’s transmission spectrum. In Cristo et al, in prep
		we will explore the impact of interpolation in transit and atmosphere characterization.
		\item Science cases that hinge on intensive observation campaigns on small time intervals (corresponding to BERV excursions comparable to the typical pixel size of the instrument) will be heavily affected by our assumptions. Asteroseismic analysis and studies of ultra-short period exoplanets will be particularly affected by the effects shown on this paper;
		\item The detailed characterization of planetary systems over long timespans, with the goal of detecting of Earth-like exoplanets, can also be affected with interpolation-induced signals. We estimate the upper ceiling of the RV contamination at 20 \centimetersecond{}, larger than the semi-amplitude of Earth-like planets orbiting Sun-like stars, which could impact surveys targeting such systems \citep[e.g., the Terra Hunting Experiment][]{hall_feasibility_2018}. 
		\item The direct measurement of cosmological redshift drift is one of the flagship goals of ultra-stable spectrographs such as ESPRESSO and 
		future ELT-ANDES \citep{marques_fundamental_2023,trost_espresso_2025}. Although photon-starved, the measurement is fundamentally limited by systematic effects, emphasizing the need for extreme instrumental stability and precise control over all the reduction and analysis steps. In particular, within each observation epoch, individual quasars exposures are typically resampled onto a common wavelength grid before combination, an interpolation-dependent process that must remain stable at cm/s level over decades. Interpolation induced biases at 5-20 cm/s
		level, as identified here, can dominate the expected cosmological redshift drift signal ($\sim$5 cm/s per decade) and may thus mimic or mask the actual cosmic acceleration.
	\end{itemize}
	
	The effects that were explored in this paper are thus extendable to science cases that hinge on the construction of high-SNR stellar models that require interpolation of stellar spectra. One possible avenue to mitigate such issues would be the construction of stellar templates without making use of interpolation to a common wavelength grid, or minimizing its contamination. One example of such avenue would be the concatenation of multiple observations, after they are normalized, creating at a super-sampled template that would decrease its $\Delta\lambda$. The consequence of such approach would be a smaller SNR per point of the template, as we would no longer stacking observations. For RV extraction purposes, one can entirely bypass such issues through the standard CCF method or even some template-free methodologies (e.g., \texttt{TILARA}, \citet{sannicolasmartinezTILARATemplateIndependentLinebyline2026}).

\begin{acknowledgements}
	The authors acknowledge the contribution from François Bouchy as the coordinator of the "Other Science" package from the ESPRESSO GTO.  The authors acknowledge Nicolas Cowan for discussions around such systematic issues. This work was funded by the European Union (ERC, FIERCE, 101052347). Views and opinions expressed are however those of the author(s) only and do not necessarily reflect those of the European Union or the European Research Council. Neither the European Union nor the granting authority can be held responsible for them. This work was also supported by FCT - Fundação para a Ciência e a Tecnologia through national funds by these grants: UIDB/04434/2020 DOI: 10.54499/UIDB/04434/2020, UIDP/04434/2020 DOI: 10.54499/UIDP/04434/2020, PTDC/FIS-AST/4862/2020, UID/04434/2025.
	CMJM acknowledges the supported by FCT - Fundação para a Ciência e Tecnologia, I.P. by project reference 2023.03984.BD and DOI identifier \url{https://doi.org/10.54499/2023.03984.BD}. TLC is supported by Funda\c c\~ao para a Ci\^encia e a Tecnologia (FCT) in the form of a work contract (\href{https://doi.org/10.54499/2023.08117.CEECIND/CP2839/CT0004}{2023.08117.CEECIND/CP2839/CT0004}).
	The INAF authors acknowledge financial support from the Italian Ministry of Education,	University, and Research through PRIN 201278X4FL and the ‘Progetti Premiali’ funding scheme. D.D. is supported by the Fonds de recherche du Québec—Nature et technologies (FRQNT). KA acknowledges support from the Swiss National Science Foundation (SNSF) under the Postdoc Mobility grant P500PT\_230225. 
	EA acknowledge the financial support of the FRQ-NT through the Centre de recherche en astrophysique du Québec as well as the support from the Trottier Family Foundation and the Trottier Institute for Research on Exoplanets.	EA acknowledges support from Canada Foundation for Innovation (CFI) program, the Université de Montréal and Université Laval, the Canada Economic Development (CED) program and the Ministere of Economy, Innovation and Energy (MEIE). S.G.S acknowledges the support from FCT through Investigador FCT contract nr. CEECIND/00826/2018 and  POPH/FSE (EC).
\end{acknowledgements}

\bibliographystyle{aa} 
\bibliography{sample} 

\begin{thebibliography}{35}
\expandafter\ifx\csname natexlab\endcsname\relax\def\natexlab#1{#1}\fi

\bibitem[{Aigrain \& Foreman-Mackey(2022)}]{aigrain_gaussian_2022}
Aigrain, S. \& Foreman-Mackey, D. 2022, Gaussian {Process} regression for
  astronomical time-series, arXiv:2209.08940 [astro-ph]

\bibitem[{Al~Moulla(2025)}]{al_moulla_arve_2025}
Al~Moulla, K. 2025, A\&A, 701, A266

\bibitem[{Anglada-Escudé \& Butler(2012)}]{anglada_escude_HARPS_TERRA_2012}
Anglada-Escudé, G. \& Butler, R.~P. 2012, ApJS, 200, 15

\bibitem[{Artigau {et~al.}(2022)Artigau, Cadieux, Cook, Doyon, Vandal, Donati,
  Moutou, Delfosse, Fouqué, Martioli, Bouchy, Parsons, Carmona, Dumusque,
  Astudillo-Defru, Bonfils, \& Mignon}]{artigau_line-by-line_2022}
Artigau, E., Cadieux, C., Cook, N.~J., {et~al.} 2022, AJ, 164, 84

\bibitem[{Astudillo-Defru(2015)}]{astudillo-defru_search_2015}
Astudillo-Defru, N. 2015, {PhD}, Université Grenoble, Alpes

\bibitem[{Baranne {et~al.}(1996)Baranne, Queloz, Mayor, Adrianzyk, Knispel,
  Kohler, Lacroix, Meunier, Rimbaud, \& Vin}]{baranne_elodie_1996}
Baranne, A., Queloz, D., Mayor, M., {et~al.} 1996, Astron. Astrophys. Suppl.
  Ser., 119, 373

\bibitem[{Bauer {et~al.}(2020)Bauer, Zechmeister, Kaminski, Rodríguez~López,
  Caballero, Azzaro, Stahl, Kossakowski, Quirrenbach, Becerril~Jarque,
  Rodríguez, Amado, Seifert, Reiners, Schäfer, Ribas, Béjar,
  Cortés-Contreras, Dreizler, Hatzes, Henning, Jeffers, Kürster, Lafarga,
  Montes, Morales, Schmitt, Schweitzer, \& Solano}]{bauer_carmenes_2020}
Bauer, F.~F., Zechmeister, M., Kaminski, A., {et~al.} 2020, A\&A, 640, A50

\bibitem[{Cameron {et~al.}(2020)Cameron, Ford, Shahaf, Aigrain, Dumusque,
  Haywood, Mortier, Phillips, Buchhave, Cecconi, Cegla, Cosentino, Cretignier,
  Ghedina, Gonzalez, Latham, Lodi, Lopez-Morales, Micela, Molinari, Pepe,
  Piotto, Poretti, Queloz, Juan, Segransan, Sozzetti, Szentgyorgyi, Thompson,
  Udry, \& Watson}]{cameron_separating_2020}
Cameron, A.~C., Ford, E.~B., Shahaf, S., {et~al.} 2020, arXiv:2011.00018
  [astro-ph], arXiv: 2011.00018

\bibitem[{Campante {et~al.}(2024)Campante, Kjeldsen, Li, Lund, Silva, Corsaro,
  Gomes Da~Silva, Martins, Adibekyan, Azevedo~Silva, Bedding, Bossini, Buzasi,
  Chaplin, Costa, Cunha, Cristo, Faria, García, Huber, Lundkvist, Metcalfe,
  Monteiro, Neitzel, Nielsen, Poretti, Santos, \&
  Sousa}]{campante_expanding_2024}
Campante, T.~L., Kjeldsen, H., Li, Y., {et~al.} 2024, A\&A, 683, L16

\bibitem[{Damasso {et~al.}(2020)Damasso, Sozzetti, Lovis, Barros, Sousa,
  Demangeon, Faria, Lillo-Box, Cristiani, Pepe, Rebolo, Santos,
  Zapatero~Osorio, González~Hernández, Amate, Pasquini, Zerbi, Adibekyan,
  Abreu, Affolter, Alibert, Aliverti, Allart, Allende~Prieto, Álvarez, Alves,
  Avila, Baldini, Bandy, Benz, Bianco, Borsa, Bossini, Bourrier, Bouchy, Broeg,
  Cabral, Calderone, Cirami, Coelho, Conconi, Coretti, Cumani, Cupani,
  D’Odorico, Deiries, Dekker, Delabre, Di~Marcantonio, Dumusque, Ehrenreich,
  Figueira, Fragoso, Genolet, Genoni, Génova~Santos, Hughes, Iwert, Kerber,
  Knudstrup, Landoni, Lavie, Lizon, Lo~Curto, Maire, Martins, Mégevand,
  Mehner, Micela, Modigliani, Molaro, Monteiro, Monteiro, Moschetti, Mueller,
  Murphy, Nunes, Oggioni, Oliveira, Oshagh, Pallé, Pariani, Poretti, Rasilla,
  Rebordão, Redaelli, Riva, Santana~Tschudi, Santin, Santos, Ségransan,
  Schmidt, Segovia, Sosnowska, Spanò, Suárez~Mascareño, Tabernero, Tenegi,
  Udry, \& Zanutta}]{damasso_precise_2020}
Damasso, M., Sozzetti, A., Lovis, C., {et~al.} 2020, A\&A, 642, A31

\bibitem[{Doshi {et~al.}(2025)Doshi, Cowan, Artigau, Doyon, Silva, Moulla, \&
  Hezaveh}]{doshi_interpolation_2025}
Doshi, D., Cowan, N.~B., Artigau, A., {et~al.} 2025, The {Interpolation}
  {Constraint} in the {RV} {Analysis} of {M}-{Dwarfs} {Using} {Empirical}
  {Templates}, version Number: 1

\bibitem[{Dumusque(2018)}]{dumusque_measuring_2018}
Dumusque, X. 2018, A\&A, 620, A47

\bibitem[{Dumusque {et~al.}(2021)Dumusque, Cretignier, Sosnowska, Buchschacher,
  Lovis, Phillips, Pepe, Alesina, Buchhave, Burnier, Cecconi, Cegla, Cloutier,
  Collier~Cameron, Cosentino, Ghedina, González, Haywood, Latham, Lodi,
  López-Morales, Maldonado, Malavolta, Micela, Molinari, Mortier,
  Pérez~Ventura, Pinamonti, Poretti, Rice, Riverol, Riverol, San~Juan,
  Ségransan, Sozzetti, Thompson, Udry, \& Wilson}]{dumusque_three_2021}
Dumusque, X., Cretignier, M., Sosnowska, D., {et~al.} 2021, A\&A, 648, A103

\bibitem[{Faria {et~al.}(2022)Faria, Mascareño, Figueira, Silva, Damasso,
  Demangeon, Pepe, Santos, Rebolo, Cristiani, Adibekyan, Alibert, Allart,
  Barros, Cabral, D'Odorico, Di~Marcantonio, Dumusque, Ehrenreich, Hernández,
  Hara, Lillo-Box, Curto, Lovis, Martins, Mégevand, Mehner, Micela, Molaro,
  Nunes, Pallé, Poretti, Sousa, Sozzetti, Tabernero, Udry, \&
  Osorio}]{fariaCandidateShortperiodSubEarth2022}
Faria, J.~P., Mascareño, A.~S., Figueira, P., {et~al.} 2022, A\&A, 658, A115,
  arXiv: 2202.05188

\bibitem[{Faria {et~al.}(2018)Faria, Santos, Figueira, \&
  Brewer}]{faria_kima_2018}
Faria, J.~P., Santos, N.~C., Figueira, P., \& Brewer, B.~J. 2018, JOSS, 3, 487,
  arXiv: 1806.08305

\bibitem[{Farret~Jentink {et~al.}(2022)Farret~Jentink, Mortier, Snik, Dorval,
  Thompson, Navarro, \& Naylor}]{farret_jentink_aboras_2022}
Farret~Jentink, C., Mortier, A., Snik, F., {et~al.} 2022, in Ground-based and
  {Airborne} {Telescopes} {IX}, ed. H.~K. Marshall, J.~Spyromilio, \& T.~Usuda
  (Montréal, Canada: SPIE), 146

\bibitem[{Hall {et~al.}(2018)Hall, Thompson, Handley, \&
  Queloz}]{hall_feasibility_2018}
Hall, R.~D., Thompson, S.~J., Handley, W., \& Queloz, D. 2018, Monthly Notices
  of the Royal Astronomical Society, 479, 2968

\bibitem[{Jurgenson {et~al.}(2016)Jurgenson, Fischer, McCracken, Sawyer,
  Szymkowiak, Davis, Muller, \& Santoro}]{evans_expres_2016}
Jurgenson, C., Fischer, D., McCracken, T., {et~al.} 2016, in Ground-based and
  {Airborne} {Instrumentation} for {Astronomy} {VI}, ed. C.~J. Evans,
  L.~Simard, \& H.~Takami, Edinburgh, United Kingdom, 99086T

\bibitem[{Lin {et~al.}(2022)Lin, Monson, Mahadevan, Ninan, Halverson, Nitroy,
  Bender, Logsdon, Kanodia, Terrien, Roy, Luhn, Gupta, Ford, Hearty, Laher,
  Hunting, McBride, Salazar~Rivera, Rajagopal, Wolf, Robertson, Wright, Blake,
  Cañas, Lubar, McElwain, Ramsey, Schwab, \& Stefansson}]{lin_observing_2022}
Lin, A. S.~J., Monson, A., Mahadevan, S., {et~al.} 2022, AJ, 163, 184

\bibitem[{Marques {et~al.}(2023)Marques, Martins, \&
  Alves}]{marques_fundamental_2023}
Marques, C. M.~J., Martins, C. J. A.~P., \& Alves, C.~S. 2023, Monthly Notices
  of the Royal Astronomical Society, 522, 5973, publisher: OUP ADS Bibcode:
  2023MNRAS.522.5973M

\bibitem[{Pepe {et~al.}(2021)Pepe, Cristiani, Rebolo, Santos, Dekker, Cabral,
  Di~Marcantonio, Figueira, Curto, Lovis, Mayor, Mégevand, Molaro, Riva,
  Osorio, Amate, Manescau, Pasquini, Zerbi, Adibekyan, Abreu, Affolter,
  Alibert, Aliverti, Allart, Prieto, Álvarez, Alves, Avila, Baldini, Bandy,
  Barros, Benz, Bianco, Borsa, Bourrier, Bouchy, Broeg, Calderone, Cirami,
  Coelho, Conconi, Coretti, Cumani, Cupani, D'Odorico, Damasso, Deiries,
  Delabre, Demangeon, Dumusque, Ehrenreich, Faria, Fragoso, Genolet, Genoni,
  Santos, Hernández, Hughes, Iwert, Kerber, Knudstrup, Landoni, Lavie,
  Lillo-Box, Lizon, Maire, Martins, Mehner, Micela, Modigliani, Monteiro,
  Monteiro, Moschetti, Murphy, Nunes, Oggioni, Oliveira, Oshagh, Pallé,
  Pariani, Poretti, Rasilla, Rebordão, Redaelli, Tschudi, Santin, Santos,
  Ségransan, Schmidt, Segovia, Sosnowska, Sozzetti, Sousa, Spanò, Mascareño,
  Tabernero, Tenegi, Udry, \& Zanutta}]{pepeESPRESSOVLTOnsky2021}
Pepe, F., Cristiani, S., Rebolo, R., {et~al.} 2021, A\&A, 645, A96, arXiv:
  2010.00316

\bibitem[{Pepe {et~al.}(2002)Pepe, Mayor, Galland, Naef, Queloz, Santos, Udry,
  \& Burnet}]{pepe_coralie_2002}
Pepe, F., Mayor, M., Galland, F., {et~al.} 2002, A\&A, 388, 632

\bibitem[{Rubenzahl {et~al.}(2023)Rubenzahl, Halverson, Walawender, Hill,
  Howard, Brown, Ida, Tehero, Fulton, Gibson, Kassis, Smith, Wold, \&
  Payne}]{rubenzahl_staring_2023}
Rubenzahl, R.~A., Halverson, S., Walawender, J., {et~al.} 2023, PASP, 135,
  125002

\bibitem[{San Nicolas~Martinez {et~al.}(2026)San Nicolas~Martinez, Santos,
  Adibekyan, Al~Moulla, Silva, \&
  Sousa}]{sannicolasmartinezTILARATemplateIndependentLinebyline2026}
San Nicolas~Martinez, C., Santos, N.~C., Adibekyan, V., {et~al.} 2026, A\&A,
  708, A317

\bibitem[{Santos {et~al.}(2025)Santos, Cabral, Leite, Smette, Abreu, Alves,
  Martins, Monteiro, Silva, Wehbe, Arancibia, Ávila, Brillant, Cárdenas,
  Clara, Gafeira, Gaytan, Lovis, Miranda, Moreno, Oliveira, Otarola, Pepe,
  Rojas, Schmutzer, Sosnowska, Van Der~Heyden, Al~Moulla, Adibekyan, Barka,
  Barros, Branco, Cristo, Damasceno, Demangeon, Dethier, Faria, Gomes Da~Silva,
  Gonçalves, Lucero, Rodrigues, San Nicolas~Martinez, Santos, Sousa, \&
  Viana}]{santos_poet_2025}
Santos, N.~C., Cabral, A., Leite, I., {et~al.} 2025, Published in The Messenger
  vol. 194, pp. 21-25, 5 pages, artwork Size: 5 pages Medium: PDF Publisher:
  European Southern Observatory (ESO)

\bibitem[{Seifahrt {et~al.}(2022)Seifahrt, Bean, Kasper, Stürmer, Brady, Liu,
  Zechmeister, Stefansson, Montet, White, Tapia, \&
  Schwab}]{seifahrt_maroon-x_2022}
Seifahrt, A., Bean, J.~L., Kasper, D., {et~al.} 2022, in Ground-based and
  {Airborne} {Instrumentation} for {Astronomy} {IX}, ed. C.~J. Evans, J.~J.
  Bryant, \& K.~Motohara (Montréal, Canada: SPIE), 50

\bibitem[{Silva {et~al.}(2026)Silva, Faria, Santos, Sousa, Viana, \&
  Martins}]{silva_astra_2026}
Silva, A.~M., Faria, J.~P., Santos, N.~C., {et~al.} 2026, JOSS, 11, 9413

\bibitem[{Silva {et~al.}(2022)Silva, Faria, Santos, Sousa, Viana, Martins,
  Figueira, Lovis, Pepe, Cristiani, Rebolo, Allart, Cabral, Mehner, Sozzetti,
  Mascareño, Martins, Ehrenreich, Mégevand, Palle, Lo~Curto, Tabernero,
  Lillo-Box, \& al}]{silvaNovelFrameworkSemiBayesian2022}
Silva, A.~M., Faria, J.~P., Santos, N.~C., {et~al.} 2022, A\&A

\bibitem[{Silva {et~al.}(2025)Silva, Santos, Faria, Martins, Cristo, Sousa,
  Viana, Artigau, Al~Moulla, Castro-González, Folha, Figueira, Schmidt, Pepe,
  Dumusque, Demangeon, Campante, Delfosse, Wehbe, Lillo-Box, Costa~Silva,
  Rodrigues, González~Hernández, Azevedo~Silva, Cristiani, Tabernero, Palle,
  Lavie, Suárez~Mascareño, Di~Marcantonio, Cabral, Martins, Nunes, \&
  Sozzetti}]{silva_systematic_2025}
Silva, A.~M., Santos, N.~C., Faria, J.~P., {et~al.} 2025, A\&A, 700, A93

\bibitem[{Sousa {et~al.}(2015)Sousa, Santos, Adibekyan, Delgado-Mena, \&
  Israelian}]{sousa_ares_2015}
Sousa, S.~G., Santos, N.~C., Adibekyan, V., Delgado-Mena, E., \& Israelian, G.
  2015, A\&A, 577, A67

\bibitem[{Sousa {et~al.}(2007)Sousa, Santos, Israelian, Mayor, \&
  Monteiro}]{sousa_new_2007}
Sousa, S.~G., Santos, N.~C., Israelian, G., Mayor, M., \& Monteiro, M. J. P.
  F.~G. 2007, A\&A, 469, 783

\bibitem[{Trost {et~al.}(2025)Trost, Marques, Cristiani, Cupani, Di~Stefano,
  D'Odorico, Guarneri, Martins, Milaković, Pasquini, Génova~Santos, Molaro,
  Murphy, Nunes, Schmidt, Alibert, Boutsia, Calderone, González~Hernández,
  Grazian, Lo~Curto, Palle, Pepe, Porru, Santos, Sozzetti, Suárez~Mascareño,
  \& Zapatero~Osorio}]{trost_espresso_2025}
Trost, A., Marques, C. M.~J., Cristiani, S., {et~al.} 2025, Astronomy and
  Astrophysics, 699, A159, publisher: EDP ADS Bibcode: 2025A\&A...699A.159T

\bibitem[{Virtanen {et~al.}(2020)Virtanen, Gommers, Oliphant, Haberland, Reddy,
  Cournapeau, Burovski, Peterson, Weckesser, Bright, van~der Walt, Brett,
  Wilson, Millman, Mayorov, Nelson, Jones, Kern, Larson, Carey, Polat, Feng,
  Moore, VanderPlas, Laxalde, Perktold, Cimrman, Henriksen, Quintero, Harris,
  Archibald, Ribeiro, Pedregosa, van Mulbregt, {SciPy 1.0 Contributors},
  Vijaykumar, Bardelli, Rothberg, Hilboll, Kloeckner, Scopatz, Lee, Rokem,
  Woods, Fulton, Masson, Häggström, Fitzgerald, Nicholson, Hagen, Pasechnik,
  Olivetti, Martin, Wieser, Silva, Lenders, Wilhelm, Young, Price, Ingold,
  Allen, Lee, Audren, Probst, Dietrich, Silterra, Webber, Slavič, Nothman,
  Buchner, Kulick, Schönberger, de~Miranda~Cardoso, Reimer, Harrington,
  Rodríguez, Nunez-Iglesias, Kuczynski, Tritz, Thoma, Newville, Kümmerer,
  Bolingbroke, Tartre, Pak, Smith, Nowaczyk, Shebanov, Pavlyk, Brodtkorb, Lee,
  McGibbon, Feldbauer, Lewis, Tygier, Sievert, Vigna, Peterson, More, Pudlik,
  Oshima, Pingel, Robitaille, Spura, Jones, Cera, Leslie, Zito, Krauss,
  Upadhyay, Halchenko, \&
  Vázquez-Baeza}]{virtanenSciPyFundamentalAlgorithms2020}
Virtanen, P., Gommers, R., Oliphant, T.~E., {et~al.} 2020, Nat Methods, 17, 261

\bibitem[{Zechmeister {et~al.}(2018)Zechmeister, Reiners, Amado, Azzaro, Bauer,
  Béjar, Caballero, Guenther, Hagen, Jeffers, Kaminski, Kürster, Launhardt,
  Montes, Morales, Quirrenbach, Reffert, Ribas, Seifert, Tal-Or, \&
  Wolthoff}]{zechmeisterSpectrumRadialVelocity2018}
Zechmeister, M., Reiners, A., Amado, P.~J., {et~al.} 2018, A\&A, 609, A12

\bibitem[{Zhao {et~al.}(2022)Zhao, Ford, \&
  Tinney}]{zhaoFIESTAIIDisentangling2022}
Zhao, J., Ford, E.~B., \& Tinney, C.~G. 2022, ApJ, 935, 75

\end{thebibliography}

\begin{appendix}

	\section{Line-center impact of interpolation of a gaussian profile} \label{App:line_center_interpol}

		In this Appendix we showcase the impact of interpolation on the line properties of a Gaussian line. We start by constructing an analytical gaussian profile, which is then sampled at different positions within the lane, keeping the same $\Delta\ \lambda$ between points. Then, we interpolate this discrete gaussian profile to a finer grid, comparing the result aganst the original gaussian line. 

		\begin{figure}[h!]
			\centering
			\resizebox{\hsize}{!}{\includegraphics{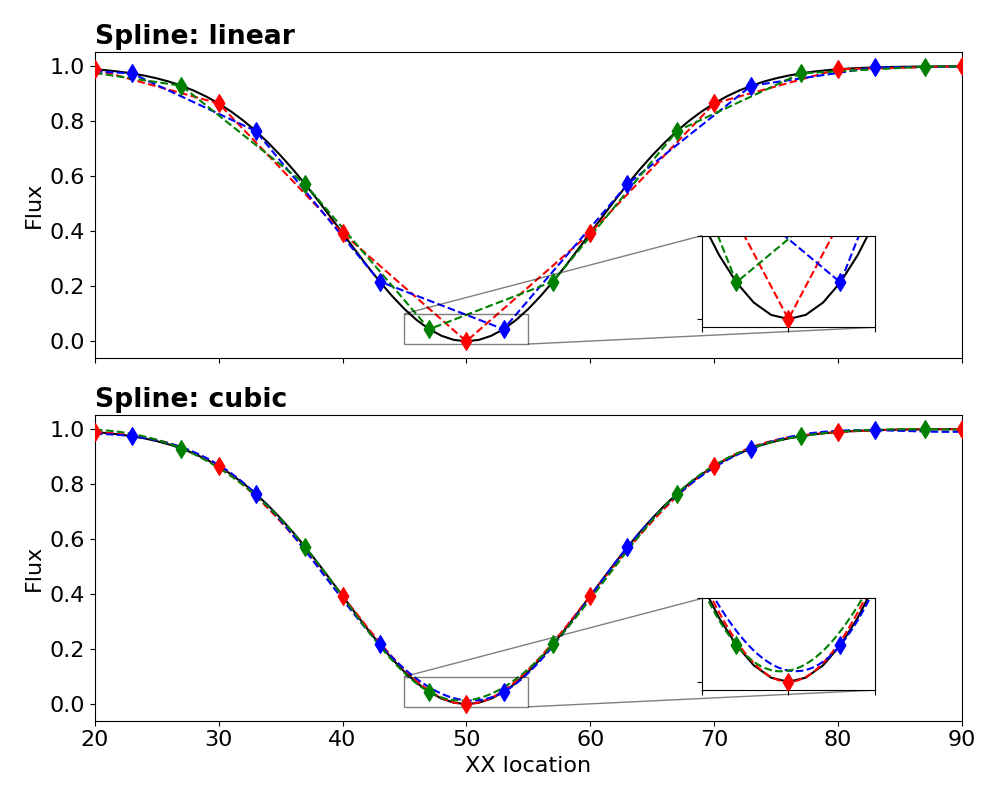}}
			\caption{Impact of the sampling location on the interpolation of a gaussian line (black line). We selected three different sub-sets of discrete locations on the gaussian profile (as shown in the blue, red, and green markers) and interpolated them to a finer wavelength grid (dashed lines). The top panel highlights the impact with a linear interpolation and the bottom one with a cubic spline interpolation.}
			\label{Fig:gaussian_showcase}
		\end{figure}

		The results of such comparison are shown in Figure \ref{Fig:gaussian_showcase}, where we find that the line depth and center are clearly affected by the choice of both interpolation algorithm and the points in which the line was sampled. As with real observations the core of the spectral line might not fall exactly in on of the wavelengths that our detector samples, an interpolation of a real spectral line will, in the majority of the cases, introduce an effect similar to the one of Figure \ref{Fig:gaussian_showcase}. Assuming this noise-free model of a spectral line, coupled with a smooth interpolation, we will introduce an apparent Doppler-shift of the line center. This can either translate into a blue-shift or a red-shift, depending on the relative position of the line center and sampling location within the detector. Furthermore, this will also introduce a decrease in the depth of the interpolated profile. In real, noisy, spectra, the impact is not as clear, as the flux variations can introduce some local sharp changes in the line profile, leading to a different pattern in the interpolated model.
	
	\section{The interplay between wavelength calibration and BERV} \label{App:single_mult_differences_linesampling} 

		One important step of the DRS extraction is the wavelength calibration, where each pixel in the CCD is associated with a given wavelength. This is done accordingly to a map constructed from a calibration frame taken with a nightly cadence. After this point, the wavelength solution is corrected from the BERV corresponding to that given moment in time. Within a single night, the BERV correction is mainly accounting for the effect of Earth's rotation, leading to a smooth variation. 
			
\begin{figure}[H]
			\centering
			\resizebox{\hsize}{!}{\includegraphics{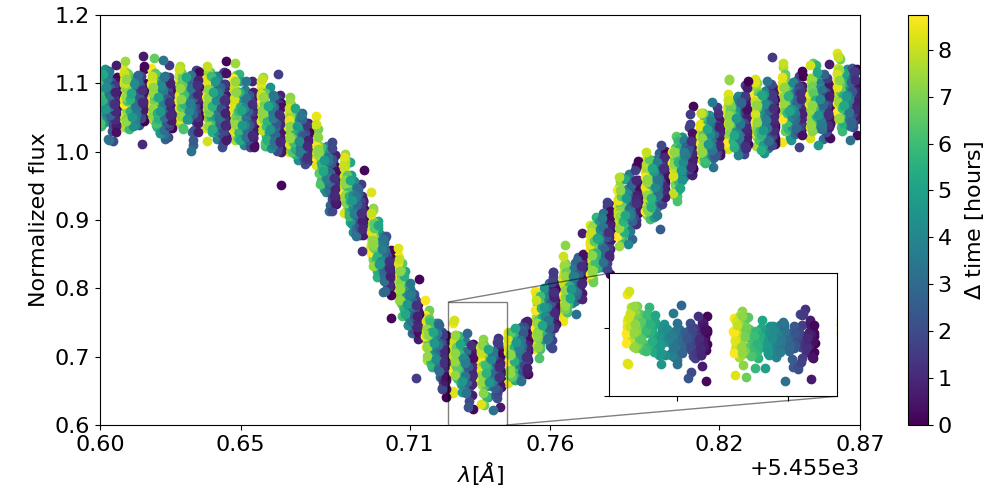}}
			\caption{Impact of the BERV on the location of one selected spectral line on the detector, over 8 hours of observations of HD40307 (the first observation of the night is represented in blue and the last in yellow).}
			\label{TREND:FIG:line_sampling_evol}
		\end{figure}

		Physically, the BERV variation during a night translates into a slow drift of the spectral lines over the detector. This physical drift will be driven by the difference in BERV between consecutive exposures which, within a single night, is at a level of the meters per second. In other words, during one single night of observations, the points in which we sample the spectral lines slowly change, as depicted in the $\sim$ 8 hours of HD40307 observations in Figure \ref{TREND:FIG:line_sampling_evol}. If we consider that the mean \espresso{} pixel size covers a RV window of $\sim$ 480 \metersecond{} and ignore any variations in stellar RV, then position of the line center will not move more than one pixel. Figure \ref{TREND:FIG:line_sampling_evol} clearly shows this variation of the position of the line center

		If we think back to the recipe of template matching algorithms, we see that we are interpolating the stellar spectra at two moments: i) when constructing the stellar template; ii) when interpolating the stellar template to the wavelength solution of the individual observations. As we are dealing with this smooth variation in wavelength (smaller than the RV-width of 1 pixel) we might need to be careful with the assumptions that we make in our interpolation algorithm. We leave for a follow-up work an in-depth analysis of the impact introduced by the assumptions of the interpolation algorithms that we use. 

		Lastly, it is also important to note that the same wavelength-calibration process is applied in the case of multi-night observations. However, as we have a larger BERV span, we will not find such a smooth evolution of the points at which the spectral lines are sampled.

				
		
		\section{Impact of the interpolation strategy on RV extraction from a single line} \label{App:single_line_int_effect}

			Throughout this manuscript we have found that the interpolation strategy introduces the deformation of the spectral line, with this profile varying for different strategies. In this Appendix we revisit the simulated dataset of Section \ref{Sec:RV_single_gauss} with different interpolation algorithms for the construction of the stellar template and subsequent comparison to the individual observations.
			 
\begin{figure}[H]
				\centering
				\resizebox{\hsize}{!}{\includegraphics{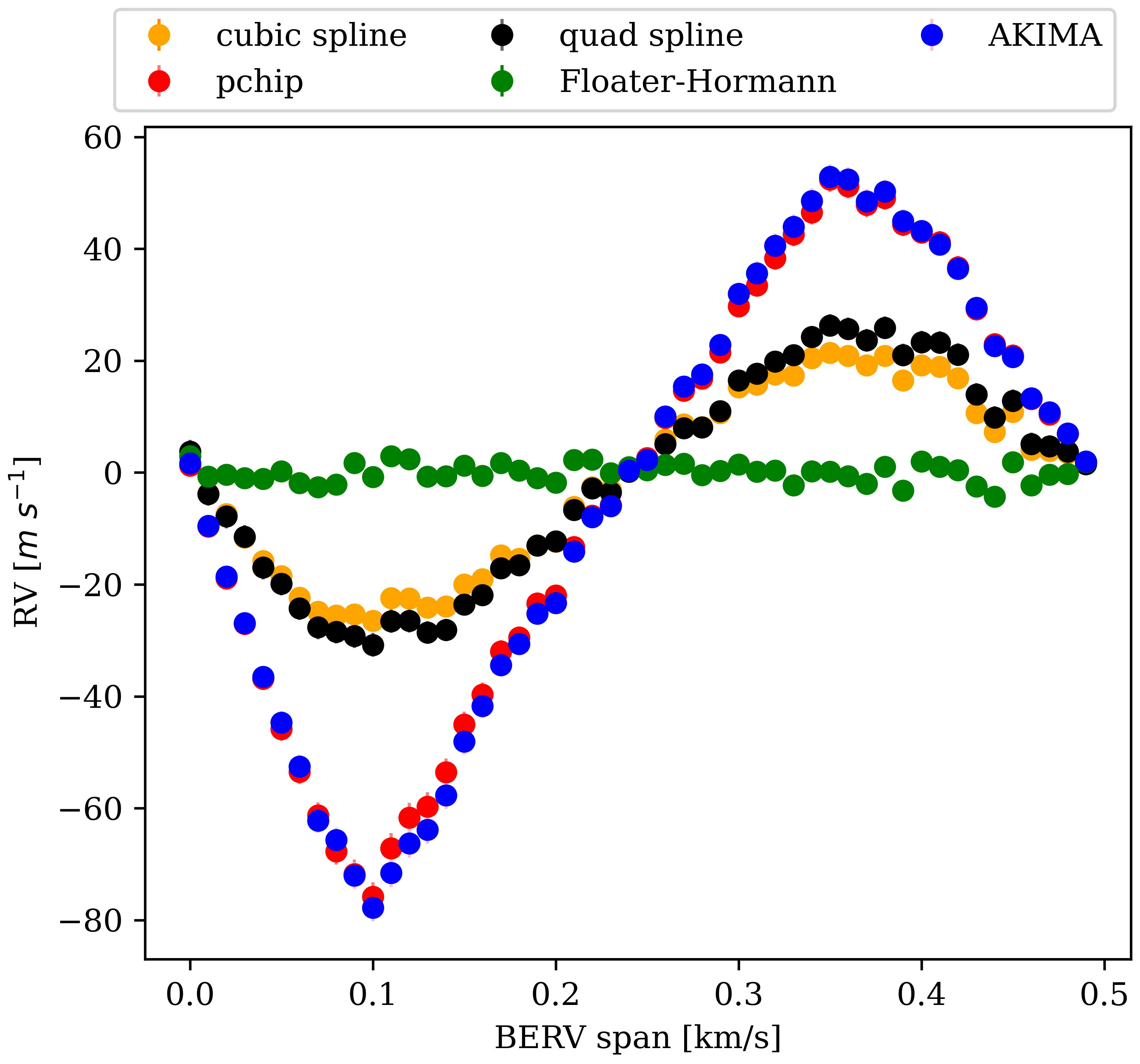}}
				\caption{Impact of the interpolation strategy on the retrieved RVs on the simulated, single gaussian line, dataset of Section \ref{Sec:RV_single_gauss}, as a function of the BERV coverage.}
				\label{TREND:FIG:single_gaussian_rv_int_effect}
			\end{figure}

			The results of this comparison are shown in \ref{TREND:FIG:single_gaussian_rv_int_effect}, where we find a multi-\metersecond{} difference amongst the different interpolation strategies explored through this paper. The larger RV effects are seen, as expected, in the methods that yield the larger flux residuals (refer to Section \ref{App:line_interpol_flux_residuals}). 

			The two methods with the smaller flux residuals (cubic spline and Floater-Hormann) reveal the smaller RV impact, albeit with a significant difference between the two. The Floater-Hormann, unlike all other interpolants reveals an almost flat profile, with a much smaller peak-to-peak amplitude. However, the analysis with real observations (Section \ref{Sec:results_intra}) reveals that this method increases the RV systematics that we measure, in comparison with the other methods. As such, we posit that this interpolator has a better performance on the simulated dataset due to the symmetrical nature of the spectral lines, which is not present on real stellar spectra.

		\subsection{Impact of spectral resolution} \label{Sect:spectral_resolution}

			Through this manuscript we have found a clear RV systematic signal that is introduced by the interpolation strategy, with a strong chromatic component. From here follows that the instrumental resolution of our spectrograph will play a critical role in the impact of such effects. In this Section we explore the dependency of the systematic signal on the instrumental resolution. For this purpose, we revisit the simulations from Section \ref{Sec:RV_single_gauss}, varying the wavelength grid of our simulations. The new wavelength grid is defined as follows:

			\begin{enumerate}
				\item From our "basis" \espresso{} observations we collect the first wavelength ($\lambda_0^{order}$) measured in each spectral order, which will be used as the start of the new wavelength grid of the corresponding order.
				\item We define a given instrumental resolution, R, which is defined as $\lambda\ /\ \Delta\lambda$.
				\item For each order, we define the wavelength grid of each point, $\lambda_i$, such that $\lambda_i\ =\ \lambda_{i-1}\ \cdot\ (1 + 1/R) $
			\end{enumerate}

			This new grid allows us to probe the impact of the instrumental resolution on the retrieved RV systematic signal, by generating noise-free datasets through the process detailed in Section \ref{Sec:RV_single_gauss}. It is important to note that this assumes a single pixel per resolution element, unlike current spectrographs where multiple CCD pixels exist per resolution element of the spectrograph. This translated into a larger $\Delta\lambda$ in these simulations, in comparison to the one that would be found on a real spectrograph with a similar instrumental resolution.

\begin{figure}[H]
				\centering
				\resizebox{\hsize}{!}{\includegraphics{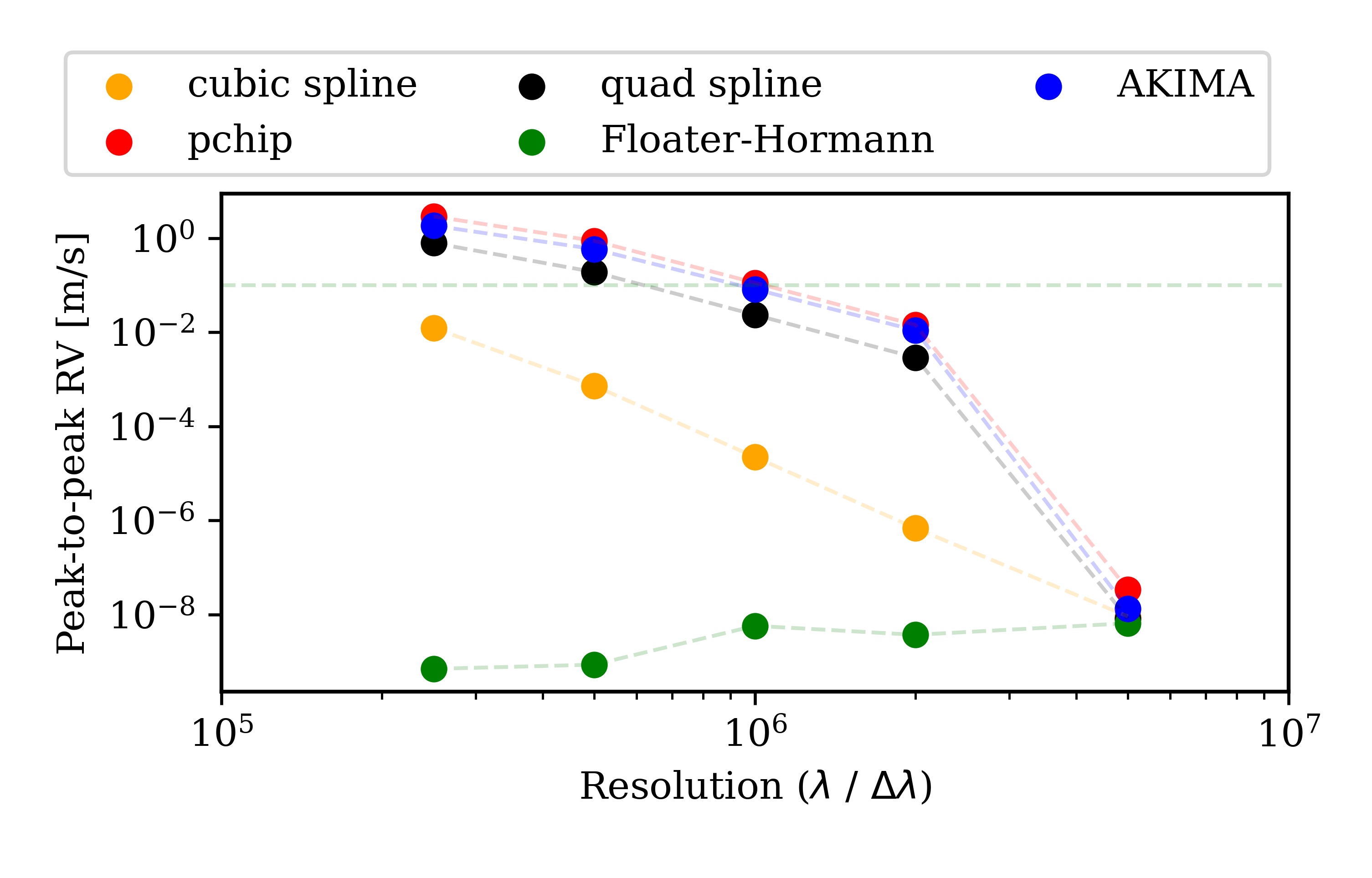}}
				\caption{Impact of the instrumental resolution on the retrieved RV systematics through different interpolation strategies. This test was done on a dataset consisting on a single, noise free, gaussian line per spectral order. The dashed green line represents the RV precision of \espresso{}.}
				\label{Fig:resolution_effect}
			\end{figure}

			In Figure \ref{Fig:resolution_effect} we find the result of this comparison, employing the different interpolation strategies explored in this paper. 

			As one would expect, in smaller instrumental resolutions (larger $\Delta\lambda$ for the same $\lambda$) the RV effect is magnified. As the resolution increases, we find that the effect decreases rapidly, falling under the noise floor of the current generation of instruments. Similarly to other tests shown in this paper, the Floater-Hormann interpolator leads to the smaller residuals, yielding a RV variation almost constant as a function of instrumental resolution. However, its performance in real observations leads to an increased systematic signal. We postulate that the blend of N-polynomials allows us to more closely capture the intrinsic shape of the spectral line in the noise-free case. In the noisier datasets, e.g., the real observations, this no longer holds true and leads to larger biases.


	\end{appendix}	
	
\end{document}